\documentclass[11pt]{article}

\usepackage{amsmath}
\usepackage{graphicx}
\usepackage{authblk}
\usepackage{algorithm}
\usepackage{amsthm}
\usepackage{amsfonts}
\usepackage{amssymb}
\usepackage{algpseudocode}
\usepackage{bbm}
\usepackage{bm}
\usepackage{booktabs}
\usepackage{caption}
\usepackage{color}
\usepackage{enumerate}
\usepackage{enumitem}
\usepackage{float}
\usepackage{mathtools} 
\usepackage{hyperref}
\usepackage{natbib}
\usepackage{comment}
\usepackage{longtable}
\usepackage{rotating}
\usepackage{xcolor}
\usepackage{fullpage}

\newtheorem{proposition}{Proposition}
\newtheorem{theorem}{Theorem}
\theoremstyle{definition}

\newtheorem{assumption}{Assumption}

\newcommand{\bbE}{\mathbb{E}}
\newcommand{\cH}{\mathcal{H}}

\newcommand{\cN}{\mathcal{N}}

\newcommand{\cG}{\mathcal{G}}
\newcommand{\cF}{\mathcal{F}}
\newcommand{\bfp}{\mathbf{p}}
\newcommand{\bfx}{\mathbf{x}}
\newcommand{\bfX}{\mathbf{X}}
\newcommand{\bfY}{\mathbf{Y}}
\newcommand{\bfW}{\mathbf{W}}
\newcommand{\bfe}{\mathbf{e}}
\newcommand{\bsdelta}{\boldsymbol{\delta}}

\newcommand{\bsbeta}{\boldsymbol{\beta}}
\newcommand{\ind}{\mathbbm{1}}
\newcommand{\indep}{\mathrel{\perp\!\!\!\perp}}

\newcommand{\up}{\mathrm{up}}
\newcommand{\BH}{\mathrm{BH}}
\newcommand{\BC}{\mathrm{BC}}
\newcommand{\ebh}{\mathrm{eBH}}

\newcommand{\fbc}{\mathrm{FBC}}
\newcommand{\fdp}{\mathrm{FDP}}
\newcommand{\fdr}{\mathrm{FDR}}
\newcommand{\lfdr}{\mathrm{Lfdr}}

\algrenewcommand\algorithmicrequire{\textbf{Input:}}
\algrenewcommand\algorithmicensure{\textbf{Output:}}
\DeclareMathOperator*{\argmin}{arg\,min}

\begin{document}
\title{A General Framework for Multiple Testing via E-value Aggregation and Data-Dependent Weighting}
\setcounter{footnote}{1}
\author{Guanxun Li$^1$, Xianyang Zhang$^{2,}$\footnote{Corresponding author: zhangxiany@stat.tamu.edu} \smallskip \\
$^1$ Department of Statistics, Beijing Normal University at Zhuhai   \\
$^2$ Department of Statistics, Texas A\&M University}
\date{} 

\maketitle

\begin{abstract}
Motivated by recent findings in \citet{li2025note}, which established an equivalence between certain p-value-based multiple testing procedures and the e-Benjamini-Hochberg procedure \citep{wang2022false}, we introduce a general framework for constructing novel multiple testing methods through the aggregation and combination of e-values. Specifically, we propose methodologies for three distinct scenarios: (i) assembly of e-values obtained from different subsets of data, simultaneously controlling group-wise and overall false discovery rates; (ii) aggregation of e-values derived from different procedures or the same procedure employing different test statistics; and (iii) adaptive multiple testing methods that incorporate external structural information to enhance statistical power. A notable feature of our approach is the use of data-dependent weighting of e-values, significantly improving the efficiency of the resulting e-Benjamini-Hochberg procedures. The construction of these weights is non-trivial and inspired by leave-one-out analysis, a widely utilized technique for proving false discovery rate control in p-value-based methodologies. We theoretically establish that the proposed e-Benjamini-Hochberg procedures, when equipped with data-dependent weights, guarantee finite-sample false discovery rate control across all three considered applications. Additionally, numerical studies illustrate the efficacy and advantages of the proposed methods within each application scenario.
\end{abstract}

\noindent%
\vfill

\newpage
\section{Introduction}
In modern scientific research involving high-dimensional data, multiple testing frequently arises as a fundamental challenge. This occurs when simultaneously evaluating a large number of hypotheses to identify significant signals, necessitating careful control of error rates such as the false discovery rate (FDR) to ensure statistical validity.

The Benjamini-Hochberg (BH) procedure \citep{benjamini1995controlling} and the Barber-Cand\`es (BC) procedure \citep{barber2015controlling} are the most commonly employed methods for controlling the FDR using p-values. Recently, there has been increasing interest in employing e-values for FDR control; see, for example, \citet{ignatiadis2024compound, ignatiadis2024values,  xu2024online}.
% \citet{xu2021unified, banerjee2023harnessing, xu2023more, ignatiadis2024values, ignatiadis2024compound, ren2024derandomised, xu2024online, fischer2024online}.
In particular, \citet{wang2022false} proposed a multiple testing approach called the e-Benjamini-Hochberg (e-BH) procedure, which applies the BH procedure directly to e-values. They demonstrated that the e-BH procedure controls the FDR even when e-values exhibit arbitrary dependence structures. Compared to p-values, which are defined through tail probabilities, e-values, defined via expectation, offer greater flexibility for combining multiple e-values to obtain new e-values \citep{vovk2021values}. For comprehensive reviews of methods based on e-values, we refer readers to \citet{ramdas2024hypothesis}.

In recent work, \citet{li2025note} introduced a unified framework for multiple testing procedures based on p-values, which includes the BH and BC procedures as special cases. The authors established the equivalence between these p-value-based methods and the e-BH procedure when appropriate sets of e-values are utilized. Here, equivalence means that these methods yield identical rejection sets. Motivated by these findings, we propose new multiple testing procedures that aggregate e-values derived from different methods or the same method with different test statistics, or combine e-values obtained from different subsets of data. Specifically, we explore three concrete scenarios, which are detailed in the subsequent sections.

In our first scenario, we consider the setting with $L$ sets of e-values derived from $L$ distinct datasets. Our goal is to aggregate these e-values into a single vector to incorporate all group-level information. We propose a procedure designed to control both the overall FDR and the group-wise FDR simultaneously.

A key application involves high-stakes decision-making, where we test $n$ hypotheses partitioned into $G$ groups according to certain protected attributes. In a loan-approval system, for example, customers may be grouped by gender or race; the null hypothesis states that a given loan should be approved. Management must control both the overall FDR—so that too few loans are rejected—and the group-wise FDR—so that no group is unfairly treated. Conventional solutions fail: applying the FDR-controlling method to all $n$ tests fails to control FDR for certain groups; testing each group at level $\alpha$ fails to control overall FDR; and a Bonferroni adjustment to $\alpha/G$ is too conservative. To address this challenge, we propose a multiple-testing procedure that uses e-values as a bridge to combine the results from different groups, controlling the FDR within each group and the overall FDR simultaneously. Specifically, we apply the BC procedure to each group, then assemble the resulting e-values with appropriate weights to form a unified e-value vector, which we pass to the e-BH procedure. We show that the resulting method simultaneously controls the FDR within each group and the overall FDR in finite samples.

In our second scenario, we consider a setting with $L$ distinct sets of e-values, which we aim to aggregate into a single e-value vector to incorporate all available information. By effectively combining e-values from diverse sources, the proposed approach enables the integration of multiple results while rigorously controlling the overall FDR. This general scenario has several specific applications. 

The first specific application is the development of a robust and efficient knockoff method capable of accommodating various underlying relationships between the response variable and predictors. In the knockoff methodology \citep{barber2015controlling}, several test statistics can be employed: Lasso-based methods excel for (near-)linear models, whereas random forest-based methods suit nonlinear settings \citep{candes2018panning}. Because the true form of the dependence is rarely known, we develop a robust and efficient knockoff procedure that can leverage the strengths of different test statistics simultaneously. Specifically, we construct a unified set of e-values by aggregating the e-values derived from knockoff methods based on various test statistics. Subsequently, we input these aggregated e-values into the e-BH procedure. We demonstrate that the resulting hybrid knockoff procedure effectively controls the FDR and maintains robust performance regardless of the true underlying relationship between the response variable and predictors.

Our second application involves developing a robust and efficient multiple-testing procedure by combining the strengths of the BH and BC procedures. The BH and BC procedures use different strategies for estimating the number of false rejections, leading to distinct performance characteristics depending on signal density and strength. In real-world applications, reporting results from the better-performing method can lead to inflated FDR and is considered a form of data snooping. To address this issue, we propose a hybrid approach that employs e-values as a bridge to integrate results from both the BH and BC procedures. To this end, we construct a unified set of e-values by suitably weighting the e-values derived from the BH and BC procedures. These combined e-values are then input into the e-BH procedure. We show that the resulting hybrid procedure maintains rigorous FDR control and can significantly improve performance relative to the weaker individual method in finite-sample settings.

In the final scenario, we consider the problem of multiple testing with external structural information in the form of covariates, which has received significant recent attention, as leveraging auxiliary information can enhance the power and interpretability of multiple-testing results in many scientific applications. Typical covariates include (i) total read counts in RNA-seq, which modulate gene-level power, and (ii) phylogenetic distances in microbiome studies, where related species share abundance patterns. A growing list of works has reflected the importance of this research direction in recent years—for instance, \citet{hu2010false, ignatiadis2016data, lei2018adapt, li2019multiple, ignatiadis2021covariate, zhang2022covariate, zhao2024tau}. However, these existing works suffer from various limitations. For example, the local FDR-based methods \citep{sun2015false, cao2022optimal} lack finite-sample FDR control and only guarantee FDR control asymptotically. The weighted BH methods \citep{ignatiadis2021covariate, li2019multiple} lead to suboptimal power, as observed in our numerical studies. To address these drawbacks, we propose a powerful multiple-testing procedure that incorporates auxiliary information while guaranteeing FDR control in finite samples. Specifically, we randomly split the data into several disjoint groups and use a cross-fitting approach \citep{ignatiadis2021covariate} to estimate the rejection function for each group using all samples in the other groups. We then apply the flexible BC procedure \citep{li2025note} within each group, assemble the resulting group-wise BC e-values with appropriate weights, and feed the combined vector into the e-BH procedure. We show that the proposed method controls the FDR at the desired level in finite samples and achieves competitive power relative to state-of-the-art methods.

Our approach involves a data-dependent method for weighting e-values when aggregating them from the BH and BC procedures (the second scenario) or assembling them from different subsets of the data (the first and third scenarios). It is important to note that our weighting method differs from the ``boosting factor'' proposed by \citet{wang2022false}, which involves multiplying each e-value by a factor to boost them up before applying the e-BH procedure. It is also worth mentioning the connection to the work of \citet{ignatiadis2024values}, where the authors used e-values as unnormalized weights for p-values in order to improve the testing power. The authors constructed these e-values using Basu's theorem, which makes their weights independent of the p-values. In contrast, our weights are dependent on the e-values. The construction of our weights is motivated by the leave-one-out analysis for the BH and BC procedures. It ensures that the weighted e-values satisfy Condition \eqref{eq:evalue} below, which is sufficient for the corresponding e-BH procedure to maintain FDR control at the desired level. Our numerical findings show that implementing the e-BH procedure with data-dependent weights improves its efficiency across all applications compared to the unweighted version. 

The remainder of the paper is organized as follows. Section~\ref{sec:prelim} briefly reviews several multiple testing procedures and their relationship to the e-BH procedure. Sections~\ref{sec:assm}-\ref{sec:stru-test} respectively present three distinct scenarios along with our newly proposed multiple testing methodologies tailored to different applications: (i) a multiple testing procedure controlling both group-wise and overall FDR, (ii) a hybrid knockoff method and a hybrid approach that integrates the BH and BC procedures, and (iii) a structure-adaptive multiple testing procedure. Section~\ref{sec:con} provides concluding remarks. The Supplement includes additional numerical results as well as complete proofs of all main theoretical results.

\section{Preliminaries}\label{sec:prelim}
In recent work, \citet{li2025note} introduced a unified framework for understanding many commonly used multiple testing procedures and demonstrated their equivalence to the e-BH procedure using appropriately defined sets of e-values. In this section, we briefly review their results.

Consider \(n\) hypotheses \(H_1,\dots,H_n\). Let \(\mathcal H_0\) and \(\mathcal H_1\) denote the sets of true null and true alternative hypotheses, respectively. Let $\boldsymbol{\theta} = (\theta_1, \dots, \theta_n) \in \{0, 1\}^n$ represent the true states of these hypotheses, where $\theta_i = 0$ indicates that $H_i$ is under the null and $\theta_i = 1$ otherwise. We define a decision rule $\boldsymbol{\delta} = (\delta_1, \dots, \delta_n) \in \{0, 1\}^n$, where $\delta_i = 1$ indicates rejection of $H_i$, and $\delta_i = 0$ indicates acceptance.

The false discovery rate (FDR) associated with a decision rule $\boldsymbol{\delta}$ is defined as the expectation of the false discovery proportion (FDP):
\[\fdr(\bsdelta) = \bbE[\fdp(\bsdelta)],\quad \text{where}\quad \fdp(\bsdelta)=\frac{\sum_{i=1}^n (1 - \theta_i)\delta_i}{1\vee\sum_{i=1}^n \delta_i},\]
with $a\vee b = \max\{a, b\}$. An FDR-controlling procedure ensures that the FDR does not exceed a pre-specified threshold $\alpha \in (0,1)$.

\subsection{Multiple Testing Procedures}
Suppose we observe a set of p-values $p_1, p_2, \dots, p_n$ corresponding to the hypotheses $H_1, H_2, \dots, H_n$. \citet{li2025note} summarized several commonly used multiple testing procedures using the following unified form. Consider
\begin{equation}\label{eq:thres-p}
T = \sup\left\{t \in \mathcal{T} : \frac{m(t)}{1 \vee \sum_{j=1}^n R_j(t)} \leq \alpha \right\},    
\end{equation}
where $\mathcal{T}$ denotes the domain of the threshold, $m(t)$ provides a conservative estimate of the number of false rejections, and $R_i(t)$ indicates whether to reject the $i$th hypothesis at threshold $t$. The decision rule rejects hypothesis $H_i$ if and only if $R_i(T) = 1$. 

To specify an FDR-controlling procedure within this framework, one must define the functions $m(t)$ and $R_i(t)$. The most widely used method, the Benjamini-Hochberg (BH) procedure \citep{benjamini1995controlling}, sets  $m(t) = n t$ and $R_i(t) = \ind\{p_i \leq t\}$, where $\ind\{A\}$ is the indicator function associated with set $A$. Storey's (ST) procedure \citep{storey2002direct,storey2004strong} refines the BH procedure by estimating the proportion of true null hypotheses from the observed p-values. Specifically, the ST procedure defines $R_i(t) = \ind\{p_i \leq t\}$ and $m(t) = n \pi_0^{\lambda} t$, where $\pi_0^{\lambda} \coloneqq \bigl\{1 + n - R(\lambda)\bigr\} / \bigl\{(1 - \lambda)n\bigr\}$ for a fixed $\lambda \in [0, 1)$, and $R(\lambda)$ is the number of hypotheses rejected at threshold $\lambda$. \citet{barber2015controlling} introduced the Barber-Cand\`es (BC) procedure, a model-free approach leveraging symmetry properties of null p-values or test statistics to estimate false rejections. The BC procedure defines $m(t) = 1 + \sum_{i=1}^n \ind\{p_i \geq 1 - t\}$ and $R_i(t) = \ind\{p_i \leq t\}$. The flexible BC (FBC) procedure proposed by \citet{li2025note} generalizes the BC approach using hypothesis-specific rejection functions $\varphi_i$, given by $m(t) = 1 + \sum_{i=1}^n \ind\{\varphi_i(1 - p_i) \leq t\}$ and $R_i(t) = \ind\{\varphi_i(p_i) \leq t\}$. Table A.1 in Supplement A.1 summarizes the specifications of $m(t)$ and $R_i(t)$ for these procedures.

\subsection{E-Values and the e-BH Procedure}
A non-negative random variable $e$ is called an e-value if it satisfies the condition $\mathbb{E}[e] \leq 1$ under the null hypothesis. Suppose we have $n$ e-values, denoted as $e_1, e_2, \dots, e_n$, corresponding to the hypotheses $H_1, H_2, \dots, H_n$. The $\alpha$-level e-BH procedure \citep{wang2022false} involves sorting the e-values in decreasing order $e_{(1)} \geq e_{(2)} \geq \dots \geq e_{(n)},$ and rejecting the hypotheses associated with the $\hat{k}$ largest e-values, where $\hat{k} \coloneqq \max\left\{1 \leq i \leq n \colon e_{(i)} \geq n / (i\alpha)\right\}$.

Let $\mathcal{H}_0 = \{1 \leq i \leq n \colon \theta_i = 0\}$ be the set of true null hypotheses. According to Theorem 2 of \citet{wang2022false}, a key advantage of the e-BH procedure is that it controls the FDR at level $\alpha$, even when the e-values exhibit arbitrary dependence. 

\begin{proposition}[\citet{wang2022false}, Theorem~2]\label{prop:ebh-control}
\rm Suppose the set of e-values $\{e_i\}_{1 \leq i \leq n}$ satisfies
\begin{equation}\label{eq:evalue}
    \sum_{i \in \mathcal{H}_0} \mathbb{E}[e_i] \leq n.
\end{equation}
Then, the e-BH procedure controls the FDR at level $\alpha$.
\end{proposition}

In the context of multiple testing, the requirement that $\mathbb{E}[e] \leq 1$ in the definition of e-values can be relaxed. Specifically, throughout the rest of this paper, we refer to $\{e_i\}$ as a set of e-values if they satisfy Condition~\eqref{eq:evalue}.

\subsection{Connection Between Multiple-Testing Procedures and the e-BH Procedure}
Given the threshold $T$ defined in \eqref{eq:thres-p}, the e-BH procedure, defined based on the e-values $e_i = \bigl(n R_i(T)\bigr) / m(T)$ for $1 \leq i \leq n$, is equivalent to the multiple testing procedures presented in Table A.1 with the same $m(\cdot)$ and $R_i(\cdot)$ functions \citep{li2025note}. Here, equivalence means that they produce the same set of rejections.

Unlike p-values, which are defined in terms of tail probabilities, e-values are defined through expectations, making them easier to aggregate or combine. For instance, the arithmetic mean of multiple e-values remains a valid e-value. Building upon the insight that the BH and BC procedures, along with their generalized versions, are equivalent to the e-BH procedure when based on specific forms of e-values, we develop novel multiple testing methods by aggregating e-values derived from different procedures or by assembling e-values obtained from the same procedure applied to distinct subsets of data. By ensuring that these aggregated or assembled e-values satisfy Condition~\eqref{eq:evalue}, the resulting e-BH procedures maintain finite-sample FDR control. We illustrate these developments with several concrete applications in Sections \ref{sec:assm}-\ref{sec:stru-test}.

\section{Assembling E-Values Across Data Subsets}\label{sec:assm}
In this section, we consider a scenario that we have $L$ sets of e-values, $\{e_i^l: i \in \cG_l, |\cG_l| = n_l\}$, derived from $L$ distinct datasets, where $\bigcup_{l}\mathcal{G}_l = [n]$ and $\mathcal{G}_{l_1} \cap \mathcal{G}_{l_2} = \emptyset$ for $l_1 \neq l_2$. Each $e_i^l$ is associated with the hypothesis $H_i$, and $\sum_{i \in \cG_l \cap \cH_0} \mathbb{E}[e_i^l] \leq n_l$. In this context, our objective is to combine the $L$ sets of e-values into a single e-value vector $(e_1, \dots, e_n)$ that satisfies Condition~\eqref{eq:evalue}.

\subsection{Simultaneous Group-Wise and Overall FDR Control}
Recall that $\theta_i \in \{0,1\}$ represents the true state of hypothesis $H_i$, and $\delta_i \in \{0,1\}$ denotes the decision rule for $H_i$. We define the group-wise FDP and FDR based on $\bsdelta$ as follows:
\begin{align*}
\fdp_l(\bsdelta)=\frac{\sum_{i\in \cG_l}(1-\theta_i)\delta_i}{1\vee\sum_{i\in \cG_l}\delta_i},\quad \text{FDR}_l(\bsdelta)=\mathbb{E}[\text{FDP}_l(\bsdelta)], \quad l=1,2,\dots,L.   
\end{align*}
A decision rule $\bsdelta$ with target FDR level $\alpha$ is said to simultaneously control both group-wise and overall FDR if it uniformly controls the group-wise FDRs for all $1 \leq l \leq L$ and maintains the overall FDR at level $\alpha$. Specifically, this means that $\max_{1 \leq l \leq L} \text{FDR}_l(\bsdelta) \leq \alpha$ and  $\text{FDR}(\bsdelta) \leq \alpha$. 

A decision rule that simultaneously controls both group-wise and overall FDR is relevant to predictive parity within the classification context in the fairness community \citep{chouldechova2017fair}. We compare our definitions with predictive parity in Supplement D.

We propose a multiple testing procedure that simultaneously controls both group-wise and overall FDR by assembling the e-values from the BC procedure applied to each group separately. Specifically, we implement the BC procedure at the level $\alpha$ for each individual group and let 
\begin{equation}\label{eq:threshold-fairness}
T_l = \sup\left\{0 < t < 0.5\colon \frac{1 + \sum_{i\in\cG_l}\ind\{p_{i} \geq 1 - t\}}{1\vee\sum_{i\in\cG_l}\ind\{p_{i} \leq t\}}\leq \alpha \right\}
\end{equation}
be the rejection threshold for the $l$th group with $1\leq l\leq L$. Define 
\begin{equation}\label{eq:evalue-fairness}
e_{i} = \frac{n_l w_{i}\ind\{p_{i} \leq T_l\}}{1 + \sum_{j\in\cG_l}\ind\{p_{j}\geq 1 - T_l\}},
\end{equation}
for $i\in \cG_l$, where $w_{i}>0$ represents the weight for the $i$th hypothesis, which will be specified in Section~\ref{sec:fair-para}. After collecting the e-values from each group, we implement the e-BH procedure at level $\alpha$. The testing procedure is summarized in Algorithm~\ref{alg:fair}.

\begin{algorithm}  
\caption{Multiple testing procedure that simultaneously controls both group-wise and overall FDR}\label{alg:fair}
\begin{algorithmic}[1]
\Require p-values $p_1, \dots, p_n$, group indices $\cG_1, \dots, \cG_L$, significance level $\alpha$ 
\For{$l = 1, \dots, L$}
\State Implement the BC procedure utilizing the p-values $\{p_i \colon i\in \cG_l\}$ at the level $\alpha$.
\State Calculate the threshold $T_l$ using \eqref{eq:threshold-fairness}.
\For{$i \in\cG_l$}
\State Calculate the e-value $e_{i}$ using \eqref{eq:evalue-fairness}.
\EndFor
\EndFor
\State Assemble the e-values from all groups.
\State Run the e-BH procedure utilizing the assembled e-values at the level $\alpha$.
\Ensure The indices of rejected hypotheses.
\end{algorithmic}
\end{algorithm}

It is important to note that only the nonzero e-values can be rejected in the e-BH procedure. As a result, the group-wise FDR is effectively controlled at level $\alpha$ for each group. In the following section, we will demonstrate that the e-BH procedure can effectively control the overall FDR even when the weights are selected in a data-dependent manner.

\subsubsection{Choosing Weights and FDR Control}\label{sec:fair-para}
Controlling the overall FDR requires that the e-values defined in \eqref{eq:evalue-fairness} satisfy Condition \eqref{eq:evalue}. One approach is to set $w_{i} = 1$ for all $i$. Alternatively, the group size can be taken into account by setting $w_{i} = n / (L n_l)$ for all $i \in \cG_l$ and $l = 1, \dots, L$. According to Proposition 6 in \citet{li2025note}, both strategies satisfy Condition \eqref{eq:evalue}. The e-BH procedures based on these weight choices are referred to as \texttt{eBH\_1} and \texttt{eBH\_2}, respectively. However, our simulations indicate that \texttt{eBH\_1} and \texttt{eBH\_2} often suffer from low statistical power. To enhance efficiency, we propose using a data-dependent weight approach inspired by the leave-one-out technique \citep{barber2020robust}.

Denote the p-values in the $l$th group by $\bfp_l = \{p_i\}_{i\in\cG_l}$. Write $\tilde p_i = \min\{p_i, 1 - p_i\}$, and let $\bfp_{l, i}$ for $i\in\cG_l$ be the collection of p-values obtained by replacing $p_{i}$ with $\tilde p_i$ in $\bfp_l$. By viewing $T_l$ as a functions of $\bfp_l$, we define $T_{l, i} = T_l(\bfp_{l, i})$, i.e., the threshold of the BC procedure applied to the set of p-values $\bfp_{l, i}$. We define the data-dependent weights as
\begin{equation}\label{eq:fair-weight}
w_i = \frac{\frac{n}{n_l}\left(1 + \sum_{j\neq i, j\in\cG_l}\ind\{p_j \geq 1 - T_l\}\right)}{\left(1 + \sum_{j \neq i, j \in\cG_l}\ind\{p_j\geq 1 - T_l\}\right) + \sum_{l'\neq l}\sum_{j\in\cG_{l'}}\ind\{p_{j} \geq 1 - T_{l', j}\}},
\end{equation}
for $i\in\cG_l$. The e-BH procedure, based on the weights specified in \eqref{eq:fair-weight}, will henceforth be referred to as the \texttt{eBH\_Ada} method in the following discussions. If the null $p$-values satisfy the following condition:
\begin{equation}\label{eq:ass-p-bc}
P(p_i \leq a) \leq P(p_i \geq 1 - a) = P(1 - p_i \leq a),\quad \text{ for all } 0\leq a\leq 0.5,
\end{equation}
then \texttt{eBH\_Ada} has finite-sample FDR control.

\begin{theorem}\label{thm:fair-weight}
\rm Suppose that the null p-values $\{p_i\}_{i\in\cH_0}$ are mutually independent and satisfy Condition \eqref{eq:ass-p-bc}, and are independent of the alternative p-values $\{p_i\}_{i\notin\cH_0}$. Then, the e-values specified in \eqref{eq:evalue-fairness} with the weights defined via \eqref{eq:fair-weight} satisfy Condition \eqref{eq:evalue}. Hence, the corresponding e-BH procedure controls the overall FDR in finite sample.
\end{theorem}

\subsubsection{Illustrative Example of Data-Dependent Weights}
In this subsection, we present a toy example to illustrate the effectiveness of data-dependent weights. Consider a scenario where we have two groups of p-values. The first group contains \(n_1 = 100\) p-values, and the second group contains \(n_2 = 1000\) p-values. We apply the BC method with a threshold of \(0.05\). In each group, we identify \(n_{a1} = n_{a2} = 20\) significant features, yielding e-values of \(e_1 = 100\) and \(e_2 = 1000\) for the first and second groups, respectively.

Next, we combine the e-values from both groups into a single vector and sort them in decreasing order as \(e_{(1)} \geq e_{(2)} \geq \cdots \geq e_{(n)}\), where \(n = n_1 + n_2 = 1100\). To select \(40\) significant features, the combined e-values should satisfy $e_{(40)} \geq \frac{1100}{40 \times 0.05} = 550$. Similarly, to select \(20\) significant features, the combined e-values should satisfy $e_{(20)} \geq \frac{1100}{20 \times 0.05} = 1100$. Without weighting, however, we find that \(e_{(20)} = 1000 < 1100\) and \(e_{(40)} = 100 < 550\). Thus, when the e-values are assembled without weights and the e-BH method is applied, no hypotheses are identified as significant.

\begin{figure}
    \centering
    \includegraphics[width=0.75\linewidth]{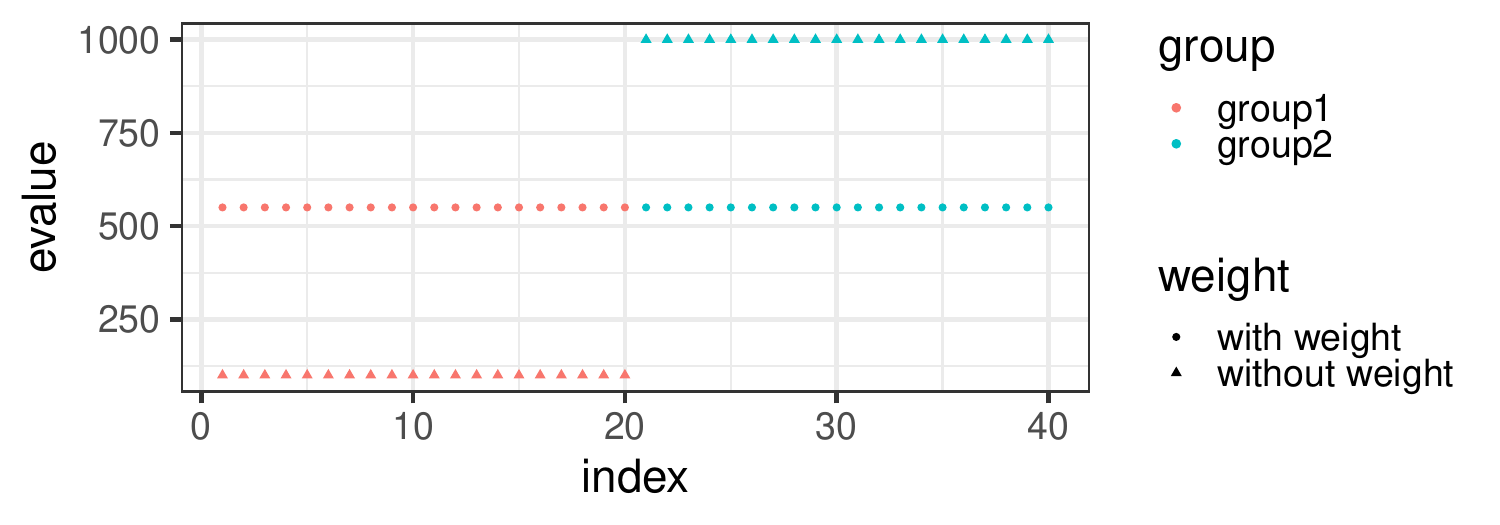}
    \caption{Comparison of weighted and unweighted e-values. Circles represent weighted e-values, while triangles denote unweighted e-values. Different colors indicate two groups.}
    \label{fig:ill-weight}
\end{figure}

Figure~\ref{fig:ill-weight} illustrates the scenario with only the non-zero e-values. Circles represent the weighted e-values, while triangles denote the e-values without weights. Different colors indicate the two groups. When e-values are combined without weights and the e-BH method is applied, no hypotheses are significant. In contrast, when the e-values are combined with weights, all hypotheses become significant. This demonstrates that data-dependent weights serve to increase the smaller e-values and decrease the larger e-values, thereby enhancing the method's ability to identify more significant discoveries while maintaining control over the FDR.

\subsubsection{Numerical Studies}\label{sec:fair-examp}
We shall compare the finite sample performance of the proposed method with two naive approaches through simulations. The first method disregards the group information and directly applies the BC procedure to all p-values. We refer to this method as \texttt{BC\_Com} for future reference. \texttt{BC\_Com} has two shortcomings. Firstly, it may fail to control the group-wise FDRs, as illustrated in Setting E2. Secondly, it fails to ensure comparable power across different groups, resulting in the possibility of one group having high power. In contrast, the other group has nearly zero power, as illustrated in Setting E1. The same issue is also encountered by \texttt{eBH\_1}. An alternative approach involves implementing the BC procedure for each group separately and combining all rejections. We call this method \texttt{BC\_Sep}. Although \texttt{BC\_Sep} effectively controls the FDR for individual groups, it does not guarantee the overall FDR control. 

We first consider the case of two groups. To evaluate each method, we employ the following metrics: POW represents the overall power combining the rejections from both groups; POW$_1$ denotes the power for the first group, while POW$_2$ represents the power for the second group. Similarly, we can define FDR, FDR$_1$, and FDR$_2$. The empirical power and FDR are computed based on 1,000 independent Monte Carlo simulations. 

In all settings, we assume that the p-values follow the uniform distribution on $[0, 1]$ under the null. For the first group, the p-value is supposed to follow Beta($\alpha_1, \beta_1$) under the alternatives, while for the second group, it follows Beta($\alpha_2, \beta_2$) under the alternatives. The parameter values for different settings are detailed in Table D.1 in Supplement D.1.

Setting E1 corresponds to a scenario in which, for instance, the first group consists of ethnic minorities while the second group comprises ethnic majorities. The number of non-nulls is the same across the two groups. The alternative p-values in the first group are larger than those in the second group on average. The results for Setting E1 are presented in Table~\ref{tab:fair-exp}. \texttt{BC\_Com} exhibits high power for the second group, yet its power in the first group is quite low. This is because the non-null p-values from the first group are not sufficiently small, and a combined analysis of the two groups demands a lower threshold, which thus fails to reject them. Additionally, \texttt{BC\_Sep} has an inflated overall FDR in this case.

\begin{table}[t]
    \centering
    \footnotesize                      % \scriptsize for extra compression
    \setlength{\tabcolsep}{3pt}        % tighter column spacing for this table
    \begin{tabular}{lcccccc ccccccc}
        \toprule
        \multicolumn{1}{c}{} & 
        \multicolumn{6}{c}{\textbf{Setting E1}} & 
        \multicolumn{6}{c}{\textbf{Setting E2}} \\
        \cmidrule(lr){2-7}\cmidrule(l){8-13}
        Method 
          & POW & POW$_1$ & POW$_2$ & FDR & FDR$_1$ & FDR$_2$ 
          & POW & POW$_1$ & POW$_2$ & FDR & FDR$_1$ & FDR$_2$ \\
        \midrule
        \texttt{BC\_Com}  
          & 0.21  & 0.01  & 0.41  & 0.036 & 0.043 & 0.034  
          & 0.49  & 0.49  & 0.49  & 0.036 & 0.006 & 0.062 \\

        \texttt{BC\_Sep}  
          & 0.378 & 0.336 & 0.420 & 0.060 & 0.035 & 0.035  
          & 0.416 & 0.727 & 0.105 & 0.056 & 0.048 & 0.017 \\

        \texttt{eBH\_1}   
          & 0.075 & 0.000 & 0.149 & 0.021 & 0.000 & 0.021  
          & 0.024 & 0.000 & 0.049 & 0.010 & 0.000 & 0.010 \\

        \texttt{eBH\_2}   
          & 0.127 & 0.128 & 0.126 & 0.012 & 0.013 & 0.010  
          & 0.079 & 0.082 & 0.077 & 0.009 & 0.005 & 0.012 \\

        \texttt{eBH\_Ada} 
          & 0.212 & 0.185 & 0.238 & 0.027 & 0.019 & 0.019  
          & 0.289 & 0.499 & 0.079 & 0.038 & 0.034 & 0.013 \\
        \bottomrule
    \end{tabular}
    \caption{\footnotesize False discovery rate (FDR) and power for Settings~E1 and~E2 (nominal FDR level = 5\%).}
    \label{tab:fair-exp}
\end{table}

% Setting E2 is the same as Setting E1 except that $\alpha_1 = \alpha_2 = 0.5$.
The results for Setting E2 are also presented in Table \ref{tab:fair-exp}. We observe that \texttt{BC\_Com} fails to control the FDR for the second group, which can be explained as follows. Due to the fact that the non-null p-values have a similar scale for both groups and the sample size of the first group is significantly smaller than that of the second group, \texttt{BC\_Com} has a higher threshold compared to the BC procedure applied only to the second group. This can result in an FDR inflation in the second group for \texttt{BC\_Com}. We also observe that \texttt{BC\_Sep} suffers from an overall FDR inflation. In contrast, all three variants of the e-BH procedure control the group-wise and overall FDRs at the desired level. \texttt{eBH\_Ada} has a much higher power than the other two e-value-based methods.

We present the results for $G = 4$ in Supplement D.1, where we consider three different scenarios (Settings F1-F3). In particular, we note that \texttt{BC\_Com} suffers from severe FDR inflation, with the empirical FDR reaching 0.318 at the 5\% target level in Setting F2. In Setting F3, \texttt{BC\_Sep} has an empirical overall FDR of 0.343, which is much higher than the 20\% target level.

To summarize, as seen in Settings E2 and F2, \texttt{BC\_Com} has no guarantee in controlling the group-wise FDR. On the other hand, \texttt{BC\_Sep} fails to control the overall FDR, as observed in all the settings, particularly Settings E2 and F3. The e-BH-based approaches provide both group-wise and overall FDR control. However, \texttt{eBH\_1} and \texttt{eBH\_2} may suffer from power loss under certain scenarios. In contrast, \texttt{eBH\_Ada} demonstrates consistent effectiveness across all settings by achieving both group-wise and overall FDR control, while maintaining reasonable power.

\subsubsection{Real-Data Example}
We illustrate the proposed method by conducting a differential abundance analysis using the microbiome dataset \texttt{cdi\_schubert}, obtained from the MicrobiomeHD repository \citep{duvallet2017microbiomehd}, originally collected in a case-control study comparing individuals with Clostridium difficile infection (CDI) to those without (nonCDI). After preprocessing, the feature table contains 2,293 operational taxonomic units (OTUs), representing bacterial taxa annotated at the phylum level. We then applied the LinDA method \citep{zhou2022linda} to identify taxa that differ between the CDI and nonCDI groups. Additional details about the dataset, the LinDA method, and implementation procedures are provided in Supplement D.2.

In this study, controlling the overall FDR ensures the reliability of global inference, whereas controlling the FDR within each phylum is essential for accurately interpreting results within biologically meaningful groups. The number of rejected taxa for each phylum is summarized in Table D.6 in Supplement D.2. The results show that for the phyla Bacteroidetes and Firmicutes, the \texttt{eBH\_Ada} method yields fewer rejections compared to the \texttt{BC\_Com} method, suggesting that \texttt{BC\_Com} might inadequately control FDR within these specific groups. Conversely, for the phylum Proteobacteria, the \texttt{eBH\_Ada} methods identify a greater number of rejections, mirroring the pattern observed in our simulation scenario E1, where the \texttt{BC\_Com} method exhibits reduced power in certain groups. Moreover, the two data-independent weighting schemes, \texttt{eBH\_1} and \texttt{eBH\_2}, produced no discoveries; consequently, we omit their results from the table. This outcome highlights the necessity of using data-dependent weights when aggregating e-values.

\section{Aggregating E-Values From Different Results} \label{sec:aggre}
In this section, we consider a scenario where we have $L$ sets of e-values, $\{e_i^{l}: i \in [n]\}_{l=1}^L$, potentially derived from $L$ distinct multiple testing procedures or the same multiple testing procedure with different test statistics or tuning parameters. Here, $\{e_i^l\}_{l=1}^L$ denotes the $L$ e-values associated with hypothesis $H_i$, and it satisfies $\sum_{i \in \mathcal{H}_0} \mathbb{E}[e_i^l] \leq n.$ Our objective is to aggregate these $L$ sets of e-values into a single e-value vector $[e_1, \dots, e_n]$ that satisfies Condition~\eqref{eq:evalue}.

We illustrate the proposed idea with a knockoff example in Supplement E, where we introduce a hybrid knockoff procedure that integrates multiple test statistics to achieve robustness across diverse modeling scenarios. Due to space constraints, a comprehensive discussion of the knockoff framework and the hybrid method is deferred to the Supplement.

\begin{algorithm}  
\caption{Hybrid Procedure}\label{alg:bc-bh}
\begin{algorithmic}[1]
\Require p-values $p_1, \dots, p_n$ and significance levels $\alpha_{\BH}$, $\alpha_{\BC}$, and $\alpha_{\ebh}$
\State Execute the BH procedure at significance level $\alpha_{\BH}$. Compute the threshold $T_{\BH}$ using \eqref{eq:thres-p} with the BH-specified $m(t)$ and $R_i(t)$ defined in Table A.1. Calculate the e-value for the BH procedure as
\[e_{\BH,i} = \frac{1}{T_{\BH}} \ind\{p_i \leq T_{\BH}\}.\]
\State Execute the BC procedure at significance level $\alpha_{\BC}$. Compute the threshold $T_{\BC}$ using \eqref{eq:thres-p} with the BC-specified $m(t)$ and $R_i(t)$ defined in Table A.1. Calculate the e-value for the BC procedure as
\[e_{\BC,i} = \frac{n \ind\{p_i \leq T_{\BC}\}}{1 + \sum_{j=1}^n \ind\{p_j \geq 1 - T_{\BC}\}}.\]
\State Compute the weighted averaged e-value using \eqref{eq:evalue-bhbc}.
\State Apply the e-BH procedure using the weighted average e-values at significance level $\alpha_{\ebh}$.
\Ensure Indices of rejected hypotheses.
\end{algorithmic}
\end{algorithm}

\subsection{Hybrid Multiple-Testing Procedure}\label{sec:bh-bc}
The second application of our proposed method leverages e-values to integrate results from both the BC and BH procedures. Empirical results in the literature indicate that neither the BH procedure nor the BC procedure consistently outperforms the other \citet{arias2017distribution}. In real-world applications, it is often impossible to determine which method will perform better. Applying both methods and reporting the results of the one that yields more rejections does not guarantee FDR control. In this section, we introduce a new multiple testing procedure that ensures finite-sample FDR control and maintains high power across a broader range of signals by leveraging the strengths of both the BH and BC procedures.

Let $e_{\BH,i}$ and $e_{\BC,i}$ denote the e-values from the BH and BC procedures (at significance levels $\alpha_{\BH}$ and $\alpha_{\BC}$, respectively) for testing the $i$th hypothesis. We define the weighted e-values as
\begin{equation}\label{eq:evalue-bhbc}
e_i = w_{\BH,i} e_{\BH,i} + w_{\BC,i} e_{\BC,i},
\end{equation}
which aggregates information from both the BH and BC procedures, where $w_{\BH,i}$ and $w_{\BC,i}$ are non-negative weights. We then apply the e-BH procedure to these aggregated e-values to obtain our rejection set. The detailed implementation is given in Algorithm~\ref{alg:bc-bh}.

\subsubsection{Choosing Significance Levels and Weights}\label{sec:bhbc-para}
We now discuss the choices of the significance levels $(\alpha_{\BH},\alpha_{\BC})$ and the weights $(w_{\BH,i},w_{\BC,i})$ in Algorithm 2, which play important roles in the hybrid procedure. 

Different from the target FDR level $\alpha_{\ebh}$, the choice of $(\alpha_{\BH},\alpha_{\BC})$ does not affect the FDR control level but instead affects the power of the hybrid procedure. Following the discussion in Section 3.2 of \citet{ren2024derandomised}, when there are $n_a$ non-nulls with extremely strong signals, we expect that $1 + \sum_{j = 1}^n \ind\{p_j \geq 1 - T_{\BC}\}\approx\bigl(\alpha_{\BC}n_a\bigr) / \bigl(1-\alpha_{\BC}\bigr)$. In a similar spirit, we expect the FDR of the BH procedure to be $\tau_0 \alpha_{\BH}$, where $\tau_0=n_0/n.$ Let $R_{\BH}$ be the number of rejections in the BH procedure. Then we have
$R_{\BH}\approx n_a/(1-\tau_0\alpha_{\BH})$ and $T_{\BH}\approx\bigl(\alpha_{\BH} n_a\bigr) /  \bigl(n(1-\tau_0\alpha_{\BH})\bigr).$ Thus the hybrid procedure will reject $H_i$ with $i\notin\mathcal{H}_0$ when
\begin{align*}
e_i\approx \frac{w_{\BC,i}n(1-\alpha_{\BC})}{\alpha_{\BC}n_a} + \frac{w_{\BH,i}n(1-\tau_0\alpha_{\BH})}{\alpha_{\BH} n_a} \geq \frac{n}{\alpha_{\ebh}n_a}.  
\end{align*}
We found that setting $\alpha_{\BC}=\alpha_{\BH}=\alpha_{\ebh}/(1+\alpha_{\ebh})$ fulfills the above constraint if $w_{\BC, i} + w_{\BH, i} \leq 1$,
and leads to good performance in our numerical studies.

Next, we consider the selection of $(w_{\BH,i}, w_{\BC,i})$, which balances the contributions from the two methods. A natural choice is to set $w_{\BH,i} = w_{\BC,i} = 0.5$ for all $i$. According to Proposition 5 in \citet{li2025note}, the e-values defined in \eqref{eq:evalue-bhbc} with these data-independent weights satisfy Condition \eqref{eq:evalue}. Therefore, by Proposition \ref{prop:ebh-control}, the corresponding e-BH procedure controls the FDR at the desired level.

Our simulations indicate that the e-BH procedure, which is based on averaged e-values, achieves only a slight improvement in power over the less powerful method between the BH and BC procedures. To address this issue, we introduce a data-dependent approach to construct weights. Our idea is partly motivated by the leave-one-out technique used in proving the FDR control for the BH \citep{ferreira2006benjamini} and BC procedures \citep{barber2020robust}. For $i = 1, \dots, n$, denote $\tilde p_i \coloneqq \min\{p_i, 1 - p_i\}$, $ \bfp_{-i} \coloneqq \{p_1, \dots, p_{i-1},\, \tilde p_i, \, p_{i+1}, \dots, p_n\} $ and $\tilde \bfp_{-i} \coloneqq \{\tilde p_1, \dots, \tilde p_{i-1},\, 0, \, \tilde p_{i+1}, \dots, \tilde p_n\}$. By viewing $T_{\BH}$ and $T_{\BC}$ as functions of the p-values, we define $T_{\BH, i} = T_{\BH}(\tilde \bfp_{-i})$ and $ T_{\BC, j} = T_{\BC}(\bfp_{-j})$. Further define $T_{\BC, j, i}$ in the same way as $T_{\BC,j}$ but with $p_i$ being replaced by 0 when $j\neq i$. We propose the following e-value weights
\begin{align}\label{eq:bcbh-weight}
\begin{split}
w_{\BH, i} &= \frac{T_{\BH, i}}{T_{\BH, i} + \frac{1}{n}\left(1 + \sum_{j\neq i}\ind\{p_j \geq 1 - T_{\BC, j, i}\}\right)},\\
w_{\BC, i} &= \frac{\frac{1}{n}\left(1 + \sum_{j\neq i}\ind\{p_j \geq 1 - T_{\BC}\}\right)}{\max_j T_{\BH, j} + \frac{1}{n}\left(1 + \sum_{j\neq i}\ind\{p_j \geq 1 - T_{\BC}\}\right)}.    
\end{split}
\end{align}
\begin{theorem}\label{thm:bcbh-weight}
\rm Suppose that the null p-values $\{p_i\}_{i\in\cH_0}$ are mutually independent and super-uniform, and are independent of the alternative p-values $\{p_i\}_{i\notin\cH_0}$. Let $w_{\BH, i}$ and $w_{\BC, i}$ be defined in \eqref{eq:bcbh-weight}. Then, the weighted average e-values specified in \eqref{eq:evalue-bhbc} with the data-dependent weights given in \eqref{eq:bcbh-weight} satisfy Condition \eqref{eq:evalue}.
\end{theorem}
As a consequence of Theorem \ref{thm:bcbh-weight}, the hybrid procedure with the data-dependent weights in \eqref{eq:bcbh-weight} provides finite-sample FDR control. 
Furthermore, in view of the proof of Theorem~\ref{thm:bcbh-weight}, the above conclusion remains true if we replace $T_{\BH, i}$ in \eqref{eq:bcbh-weight} by any (deterministic) function of $\tilde \bfp_{-i}$.

\subsubsection{Numerical Studies}\label{sec:bhbc-simu}
We investigate the finite sample performance of the hybrid procedure via several simulation examples. We set the significance level at $\alpha_{\ebh}=0.05$. For each experimental setting, the average FDP (which estimates the FDR) and the average power based on 500 Monte Carlo replications are reported. We consider two different ways to implement the hybrid procedure: (1) \texttt{eBH\_Ada}, for which the weights are calculated via \eqref{eq:bcbh-weight} and $\alpha_{\BH} = \alpha_{\BC} = \alpha_{\ebh} / (1 + \alpha_{\ebh})$; (2) \texttt{eBH\_Ave}, for which the weights are set as $w_{\BH, i} = w_{\BC, i} = 0.5$ for all $i = 1, \dots, n$ and $\alpha_{\BH} = \alpha_{\BC} = \alpha_{\ebh} / 2$. Notice that the weights defined in \eqref{eq:bcbh-weight} involve the term \(T_{\BC, j, i}\), which can be computationally expensive. To reduce the computational burden, we also consider a fast implementation of \texttt{eBH\_Ada}, referred to as \texttt{fast\_eBH\_Ada}, which uses the weights
\begin{align*}
w_{\BH, i} = \frac{T_{\BH, i}}{T_{\BH, i} + \frac{1}{n}\left(1 + \sum_{j\neq i}\ind\{p_j \geq 1 - T_{\BC, j}\}\right)}, 
\end{align*}
but otherwise, it is the same as \texttt{eBH\_Ada}. 

We generate p-values from two settings, Setting S1 and Setting S2, where the BH procedure outperforms the BC procedure in Setting S1 while the BC procedure provides significantly higher power in Setting S2. The simulation setups are deferred to Supplement F.

\begin{figure}
    \centering
    \includegraphics[width = 0.75\textwidth]{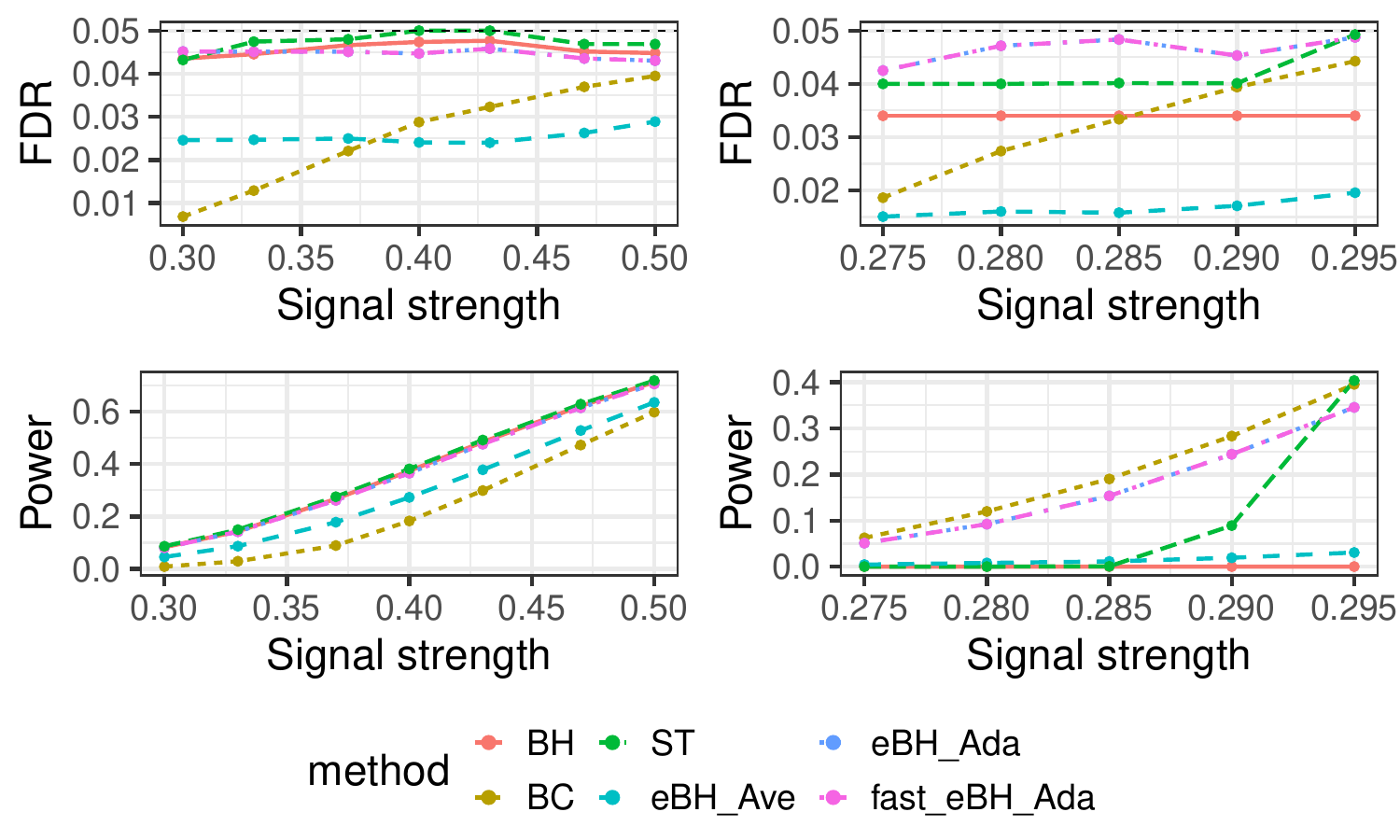}
    \caption{The left panel and right panel correspond to FDR and power for Setting S1 and Setting S2, respectively. The results are based on $500$ Monte Carlo replications.}
    \label{fig:bcbh}
\end{figure}

The left panel of Figure~\ref{fig:bcbh} summarizes the results for Setting S1. All methods under consideration control the FDR at the 5\% level. \texttt{eBH\_Ada} demonstrates nearly the same power as the BH procedure and surpasses \texttt{eBH\_Ave}, which offers a slight improvement over the BC procedure. Notably, \texttt{fast\_eBH\_Ada} achieves results almost identical to those of \texttt{eBH\_Ada} but at a significantly lower computational cost.

The right panel of Figure~\ref{fig:bcbh} summarizes the results for Setting S2. The performance of \texttt{eBH\_Ada} lies between that of the BH and BC procedures. In contrast, \texttt{eBH\_Ave} shows little improvement over the BH procedure. Again, \texttt{fast\_eBH\_Ada} produces nearly identical results to \texttt{eBH\_Ada} but with a much lower computational cost.

Next, we fix one setting in Setting S1 and calculate the computational cost for each method, as shown in Table F.1 in Supplement F. We observe that the data-independent method, \texttt{eBH\_Ave}, performs the analysis as quickly as the BH or BC procedures. The \texttt{fast\_eBH\_Ada} method demonstrates acceptable speed and is significantly faster than the \texttt{eBH\_Ada} method.

In summary, \texttt{eBH\_Ada} effectively enhances the performance of the weaker method between the BH and BC procedures across different scenarios. Notably, its power can be nearly identical to that of the stronger of the two procedures, highlighting the adaptivity of \texttt{eBH\_Ada}. \texttt{fast\_eBH\_Ada} achieves almost identical results to \texttt{eBH\_Ada} in all settings, while significantly reducing computational cost. Therefore, we recommend \texttt{fast\_eBH\_Ada} for practical applications.

\section{Structure-Adaptive Multiple Testing}\label{sec:stru-test}
Having access to various types of auxiliary information that reflect the structural relationships among hypotheses is becoming increasingly common. Taking advantage of such auxiliary information can improve the statistical power in multiple testing. In this section, we consider the scenarios where, in addition to the p-value $p_i$, there is associated structural information in the form of a covariate $x_i$ for each hypothesis. This side information represents heterogeneity among the p-values and may affect the prior probabilities of the null hypotheses being true or the signal strength under alternatives. Our goal is to develop a multiple testing procedure that can incorporate such external structural information to improve statistical power and guarantee FDR control in finite sample. The high-level idea behind our approach is to relax the p-value thresholds for hypotheses more likely to be non-null and tighten the thresholds for others through the use of a hypothesis-specific rejection rule, i.e., $\varphi_i(p_i)\leq t$, so that the FDR can be controlled.

Our proposed method combines the cross-fitting technique \citep{ignatiadis2021covariate} (a sample-splitting and fitting approach that enables learning the hypothesis-specific rejection function $\varphi_i$ without overfitting as long as the hypotheses can be partitioned into independent folds) with the FBC procedure introduced in \citet{li2025note}. First, we randomly split the data into $G$ distinct groups, denoted as $\{\cG_g \colon g = 1, \dots, G\}$, where $\bigcup_{g=1}^G \cG_g = [n]$ and $\cG_g \cap \cG_{g'} = \emptyset$ for $g\neq g'$. We estimate the rejection function $\varphi_i$ for hypothesis $i$ in the $g$th group using the data from all other groups, which ensures that the estimated rejection function is independent of the p-values in group $g$. We then apply the FBC procedure based on the estimated rejection functions separately to each group and obtain the corresponding e-values. Finally, we collect all the e-values and apply the e-BH procedure at the target level to control the FDR.

To describe the cross-fitting procedure, let us assume that $\varphi_i(p)=\varphi(p,x_i;\beta)$ for some unknown parameter $\beta$ that needs to be estimated from the data. We define the cross-fitting estimate as $\hat{\beta}_{-g} = \argmin_{\beta\in\mathcal{B}}\sum_{i\notin \mathcal{G}_g}\mathcal{L}(p_i,x_i,\beta)$. Here $\mathcal{L}$ is some loss function, such as negative log-likelihood, and $\mathcal{B}$ is a parameter space. 
Given $\hat{\beta}_{-g}$, we define $\hat{\varphi}_i(p)=\varphi(p,x_i;\hat{\beta}_{-g})$ for $i\in\mathcal{G}_g$.  Next, we apply the FBC procedure using the cross-fitted functions $\{\hat \varphi_i(\cdot)\}_{i\in\cG_g}$ at the level $\alpha_{\fbc}$. The corresponding threshold for the $g$th group is given by
\begin{equation}\label{eq:crossthres-gbc}
T_g = \sup\left\{0 < t \leq T_{g, \up} \colon \frac{1 + \sum_{i\in\cG_g}\ind\{\hat \varphi_{i}(1 - p_{i}) \leq t\}}{1\vee\sum_{i\in\cG_g} \ind\{\hat \varphi_{i}(p_{i})\leq t\}}\leq \alpha_{\fbc}\right\},
\end{equation}
where $T_{g, \up} < \min_{i\in\cG_g}\hat \varphi_i(0.5)$. We define the e-value for all $i\in\cG_g$ as
\begin{equation}\label{eq:cross-evalue-gbc}
e_{i} =\frac{n_g w_{i}\ind\{\hat \varphi_{i}(p_{i})\leq T_g\}}{1 + \sum_{j\in\cG_g}\ind\{\hat \varphi_{j}(1 - p_{j}) \leq T_g\}},
\end{equation}
where $w_{i} > 0$ represents the e-value weight for hypothesis $i$ in group $G_g$. Finally, we aggregate all e-values from each group and implement the e-BH procedure. A detailed description of our procedure is given in Algorithm G.1 in Supplement G.1. 

\subsection{Weights and FDR Control}\label{sec:weight-struc}
To ensure that the FDR is controlled at the desired level, it is crucial to verify that the e-values defined in equation \eqref{eq:cross-evalue-gbc} satisfy Condition \eqref{eq:evalue}. We shall show that under certain conditions on the weights, the e-values defined by \eqref{eq:cross-evalue-gbc} satisfy~\eqref{eq:evalue}, and as a result, the corresponding e-BH procedure controls the FDR at the desired level. Before stating the main theorem, let us first introduce some notations. Define $\tilde p_i = \min\{p_i, 1 - p_i\}$, $\bfp_g = \{p_{i}\}_{i\in\cG_g}$, and $\bfp_{g, i}$ as the collection of p-values obtained by replacing $p_{i}$ with $\tilde p_{i}$ in $\bfp_g$ for $i\in \cG_g$. 
Also, let $\bfp_{-g} = \{p_i\}_{i=1}^n \setminus \bfp_g$. Due to cross-fitting, the estimated function $\hat \varphi_{i}(\cdot)$ for $i\in\cG_g$ only depends on $\bfp_{-g}$. Moreover, given the fitted functions $\hat \varphi_{i}$ for $i\in \mathcal{G}_g$, the threshold $T_g$ defined in \eqref{eq:crossthres-gbc} can be treated as a function of $\bfp_g$. Let $T_{g, j} = T_g(\bfp_{g, j}; \{\hat \varphi_l\}_{l\in\cG_g})$.
We impose the following assumptions. 

\begin{assumption}\label{ass1}
Let $\{(p_i, x_i)\}$ for $1\leq i\leq n$ be the p-value and covariate pairs. (A) The null pairs $\{(p_i, x_i)\}_{i\in\cH_0}$ are mutually independent. (B) The null pairs $\{(p_i, x_i)\}_{i\in\cH_0}$ are independent of the alternative pairs $\{(p_i, x_i)\}_{i\notin\cH_0}$. (C) For $i\in\cH_0$, $p_i$ is independent of $x_i$ and satisfies Condition \eqref{eq:ass-p-bc}.
\end{assumption}

\begin{assumption}\label{ass2}
For all $1\leq i\leq n$, $\varphi_i(\cdot,x_i; \beta)$ is a monotonic increasing and continuous function given any $\beta$ and $x_i$.
\end{assumption}
Assumption \ref{ass1} concerns the dependence of the null pairs, which is standard in the literature; see, e.g., Assumption 1 of \citet{ignatiadis2021covariate} and  \citet{zhao2024tau}. Assumption \ref{ass2} implies that 
$P(\varphi_i(p_i,x_i;\beta) \leq b) \leq P(\varphi_i(1 - p_i,x_i;\beta) \leq b)$ for all $\varphi_i(0,x_i;\beta) \leq b \leq \varphi_i(0.5,x_i;\beta)$, which will be used in our proof. 
We shall describe a concrete choice of $\varphi_i$ in Section~\ref{sec:simu-struc} below.

\begin{theorem}\label{thm:struc-nodata}
\rm Suppose Assumptions 1 and 2 hold. If the weights $\{w_i\}$ are independent of the p-values and covariate information and satisfy 
\begin{equation}\label{eq:cond-weight-cross}
\sum_{g=1}^G n_g \max_{i\in\cG_g} w_i \leq n,
\end{equation}
then the e-values defined in \eqref{eq:cross-evalue-gbc} fulfill Condition \eqref{eq:evalue}.
\end{theorem}

A naive choice is to set $w_i = 1$ for all $i =1, 2, \dots, n$, which satisfies \eqref{eq:cond-weight-cross}. 
However, this choice of weights often leads to low statistical power in simulations. To improve efficiency, we propose a data-dependent approach for constructing the weights.
Given the group index $g$, for $g'\neq g$, $i\in\cG_g$ and $j\in\cG_{g'}$, let $\hat \varphi_j^{g, i, p}$ be the cross-fitted function obtained by replacing $\bfp_g$ with $\bfp_{g, i}(p)$, where $\bfp_{g, i}(p)$ is the collection of p-values with $p_i$ replaced by $p$. Define $T_{g', j}^{g, i, p} = T_{g'}(\bfp_{g', j};\{\hat \varphi_{l}^{g, i, p}\}_{l\in\cG_{g'}})$. For $i\in\cG_g$ with $1\leq g\leq G$, we propose the following e-value weights:
\begin{equation}\label{eq:weight-struc}
w_{i} = \frac{\frac{n}{n_g}\left(1 + \sum_{j\neq i, j\in\cG_g}\ind\{\hat \varphi_{j}(1 - p_{j})\leq T_g\}\right)}{\left(1 + \sum_{j\neq i, j\in\cG_g}\ind\{\hat \varphi_{j}(1 - p_{j})\leq T_g\}\right) + \sup_{p\in[0, 1]}\sum_{g'\neq g}\sum_{j\in\cG_{g'}}\ind\{\hat \varphi_{j}^{g, i, p}(1 - p_{j}) \leq T_{g', j}^{g, i, p}\}}. 
\end{equation}
 
The construction of $w_i$ involves taking the supremum over $p$, which is crucial for the proof to go through. On the one hand, it renders $w_i$ independent of $p_i$, which is a useful fact in the proof. On the other hand, it makes the weight sufficiently small in the sense that we can upper bound $w_i$ with the term $\sup_{p\in[0, 1]} \sum_{j\in\cG_{g'}}\ind\{\hat \varphi_{j}^{g, i, p}(1 - p_{j}) \leq T_{g', j}^{g, i, p}\}$ in its denominator replaced by $\sum_{j\in\cG_{g'}}\ind\{\hat \varphi_{j}(1 - p_{j}) \leq T_{g', j}\}$, which is another fact used in our argument.
\begin{theorem}\label{thm:struc-evalue}
\rm Suppose Assumptions 1 and 2 hold. Then, the e-values defined in \eqref{eq:cross-evalue-gbc} with weights specified by \eqref{eq:weight-struc} satisfy Condition \eqref{eq:evalue}.
\end{theorem}

The $\tau$-censored weighted BH procedure proposed by \citet{zhao2024tau} is a variant of the weighted BH procedure that employs a leave-one-out strategy to construct weights. A detailed comparison between our method and theirs is provided in Supplement G.1.

\subsection{Simulation Studies}\label{sec:simu-struc}
We shall compare the finite sample performance of the proposed method with several existing approaches through simulation studies. Throughout, we fix the sample size $n = 3,000$ and set the target FDR level at $\alpha_{\ebh} = 0.1$. For each experimental setting, we conduct 100 simulations and report the average FDP (as an estimate of the FDR) and power over the independent simulation runs.

We begin by detailing the implementation of the proposed method. In the FBC procedure, we employ a rejection rule based on the local FDR \citep{sun2007oracle} within the two-group mixture model framework. Specifically, we propose to use 
\[\varphi_i(p) = \lfdr_i(p) = \frac{\pi_i f_0(p)}{\pi_i f_0(p) + (1 - \pi_i) f_{1,i}(p)},\]
which represents the posterior probability that the $i$th hypothesis is null given the observed p-value $p$. The literature demonstrates that the rejection rule $\ind\{\varphi_i(p_i) = \lfdr_i(p_i) \leq t\}$ is optimal in maximizing the expected number of true positives among decision rules that control the marginal FDR at level $\alpha$ (see, e.g., \cite{sun2007oracle, lei2018adapt, cao2022optimal}). Additional discussions about local FDR are deferred to Supplement G.1.

Set $f_0(p)=\ind\{p\in [0,1]\}$ and $f_{1,i}(p)=(1 - \pi_i)(1-\kappa_i)p^{-\kappa_i}$ with $\kappa_i\in (0,1)$, we consider the working models proposed in \citet{zhang2022covariate}, please refer to Supplement  G.1 for more details. After getting $\hat \pi_i$ and $\hat \kappa_i$, we define the rejection rule 
\[\hat \varphi_i(p) = \frac{\hat \pi_i}{\hat{\pi}_i + (1 - \hat{\pi}_i)(1 - \hat{\kappa}_i)p^{-\hat{\kappa}_i}}\leq t.\]
We then apply the FBC procedure with the estimated $\hat{\varphi}_i$ at the target FDR level $\alpha$. The corresponding e-values are computed via \eqref{eq:cross-evalue-gbc}, and the weights are obtained from \eqref{eq:weight-struc}. To reduce computational cost, we introduce the following, less expensive weighting scheme:
\[w_{i} = \frac{\frac{n}{n_g}\left(1 + \sum_{j\neq i, j\in\cG_g}\ind\{\hat \varphi_{j}(1 - p_{j})\leq T_g\}\right)}{\left(1 + \sum_{j\neq i, j\in\cG_g}\ind\{\hat \varphi_{j}(1 - p_{j})\leq T_g\}\right) + \sum_{g'\neq g}\sum_{j\in\cG_{g'}}\ind\{\hat \varphi_{j}(1 - p_{j}) \leq T_{g', j}\}}.\] 
We refer to this method as \texttt{eBH\_FBC} for future reference. We compare the proposed method with the following competing methods: \texttt{BH}, \texttt{IHW\_storey}, \texttt{IHW\_betamix},  \texttt{AdaPT}, and \texttt{SABHA}. The implementation details of these methods are deferred to Supplement G.1. 

To illustrate the effect of the covariate, we generate a single covariate $x_i$ from the standard normal distribution. Given the value of $x_i$, we define $\pi_{i}$ as $\pi_{i} = \exp(a_0 + a_1 x_i) / \bigl(1 + \exp(a_0 + a_1 x_i)\bigr)$, where $a_0$ and $a_1$ determine the baseline signal density and the informativeness of the covariate, respectively. The values of $a_0$ and $a_1$ are fixed for each simulated dataset. Specifically, we set $a_0$ to take on values from the set $\{3.5, 2.5, 1.5\}$, achieving signal densities of approximately $3\%$, $8\%$, and $18\%$, respectively, representing sparse, medium, and dense signals. Furthermore, we set $a_1$ to take on values from the set $\{1.5, 2, 2.5\}$, representing a less informative, moderately informative, and strongly informative covariate, respectively. The underlying truth $\theta_i$ is then simulated based on $\pi_{i}:
\theta_i \sim \text{Bernoulli}(1 - \pi_{i}).$
We next generate the covariate that affects the alternative function $f_{1, i}$. Specifically, we sample another covariate $x_i' \sim \mathcal{N}(0, 1)$ and define $\eta_i = 2\exp(a_f x_i') / \bigl(1 + \exp(a_f x_i')\bigr)$, where we set $a_f \in \{0, 0.5, 1\}$ for no informativeness, less informativeness, and strong informativeness. Then, the z-scores are sampled from $z_i \sim \mathcal{N}(\eta_i\mu\theta_i, 1)$, where $\mu$ denotes the signal strength with the values evenly distributed in the interval $[2.5, 3.4]$. These $z$-scores are transformed into p-values using the one-sided formula $1 - \Phi(z_i)$. The p-values, along with the corresponding covariates $x_i$ and $x_i'$, serve as the input for the structure-adaptive multiple testing methods.

\begin{figure}
    \centering
    \includegraphics[width = 0.85\textwidth]{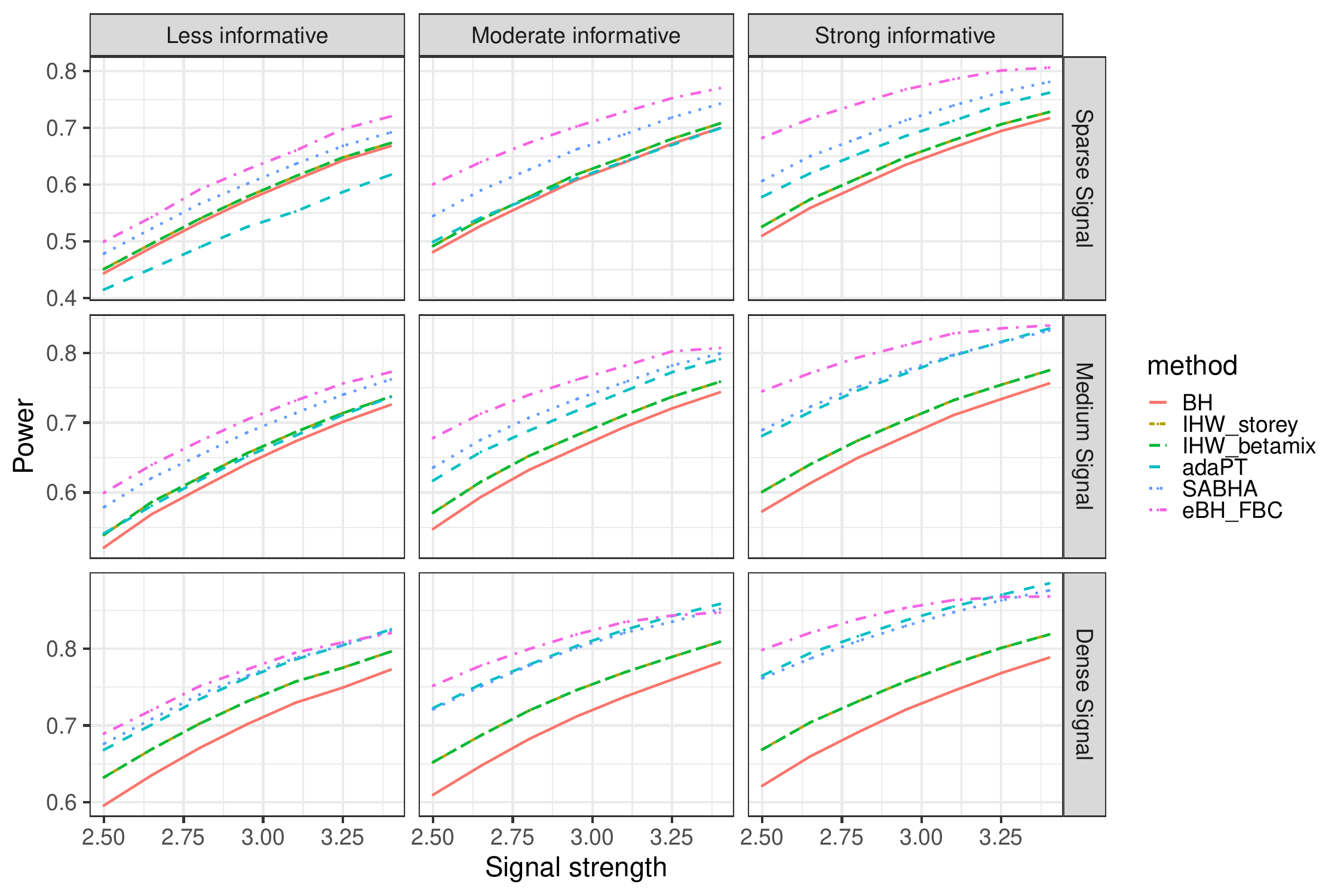}\\
    \includegraphics[width = 0.85\textwidth]{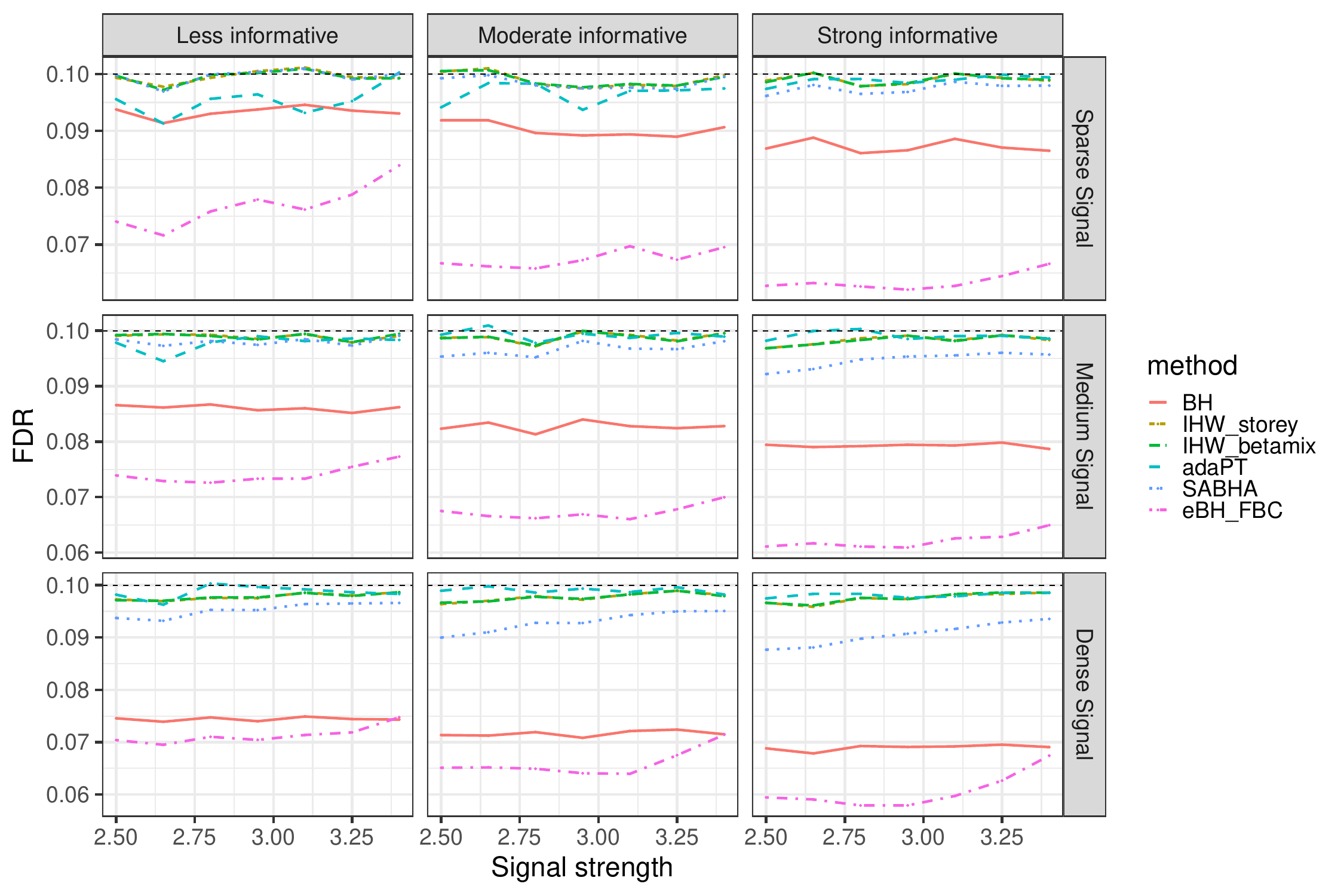}
    \caption{Empirical FDR and power with $a_f = 1$. Signal sparsity is controlled by setting $a_0\in\{3.5,2.5,1.5\}$, giving rise to sparse, moderate, and dense alternatives, respectively. Covariate informativeness is tuned via $a_1\in\{1.5,2,2.5\}$, corresponding to weak, moderate, and strong auxiliary signals.}
    \label{fig:ebh-gbc-af1}
\end{figure}

Figure~\ref{fig:ebh-gbc-af1} presents the results for $a_f = 1$. All methods successfully controlled the FDR at the desired level. When the signal is sparse ($a_0 = 3.5$), \texttt{eBH\_FBC} is the most powerful method. The SABHA method exhibits the second-best performance, while both versions of \texttt{IHW} show only slight improvements over the BH procedure. When the covariate is less informative ($a_1 = 1.5$), \texttt{AdaPT} is less powerful than the BH procedure. As the covariate becomes strongly informative ($a_1 = 2.5$), all structure-adaptive methods outperform the BH procedure in terms of power. Our proposed method demonstrates the highest power in most cases, with \texttt{AdaPT} surpassing \texttt{eBH\_FBC} in power when the signal is dense ($a_0 = 1.5$) and the covariate is strongly informative ($a_1 = 2.5$). The results for settings with $a_f = 0$ and $a_f = 0.5$ are deferred to Supplement G.2.

To demonstrate the stability of each method in structure adaptive multiple testing, particularly the consistency of results when the data are randomly generated from the same distribution, we plot the variance of the FDP in Figure G.1 in Supplement G.2. The figure shows that as the signal becomes dense or the covariate contains more information, the variance of all methods decreases. Notably, the proposed method exhibits a smaller variance compared to \texttt{AdaPT}.
Additionally, Table G.1 in Supplement G.2 displays the running time for each method to compare their computational efficiency. We focus on a simulation setting with $a_0 = 1.5$, $a_1 = 2$, $a_f = 1$, and $\mu = 3$. We conducted 100 simulation runs for each method and reported the average time taken to complete the analysis. The results indicate that our method is approximately ten times faster than \texttt{AdaPT}.

\subsubsection{Real-Data Examples}
We analyzed three omics datasets: Airway \citep{himes2014rna}, Bottomly \citep{bottomly2011evaluating}, and MWAS \citep{mcdonald2018american}. The Airway and Bottomly datasets are transcriptomics data obtained from RNA-seq experiments. For both datasets, we used the logarithm of the `basemean'' as the covariate and removed the samples with missing values, leaving us with 18,028 and 11,709 tests, respectively. We obtained the MWAS dataset from the publicly available data of the American Gut project \citep{mcdonald2018american}. We focused on a subset of subjects with ages greater than thirteen and with complete sex and country information. We excluded OTUs observed in fewer than ten subjects, resulting in 3,394 OTUs tested using the Wilcoxon rank sum test on normalized abundances. We used the library size of samples as the external covariate.

\begin{figure}
    \centering
    \includegraphics[width = 0.3\textwidth]{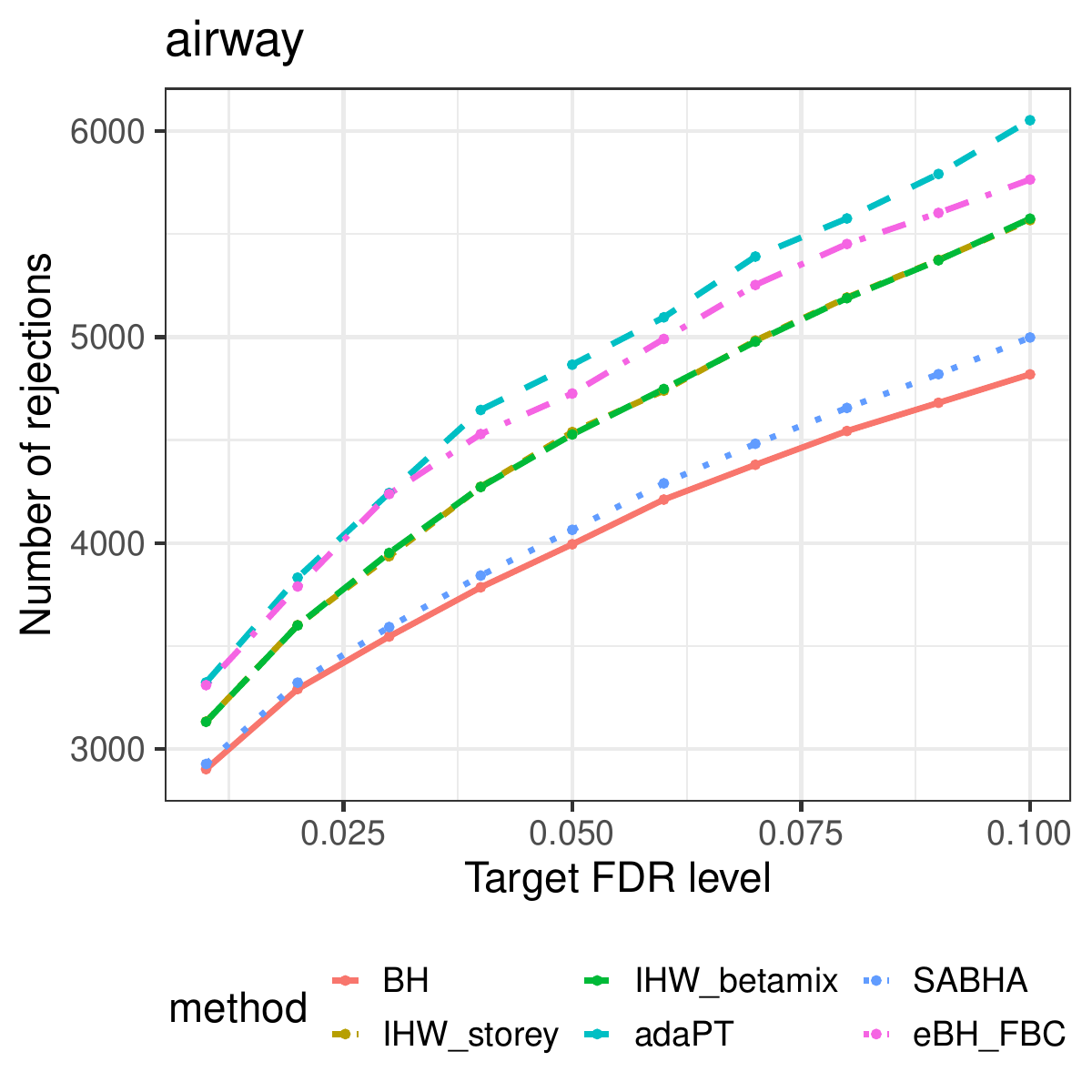}
    \includegraphics[width = 0.3\textwidth]{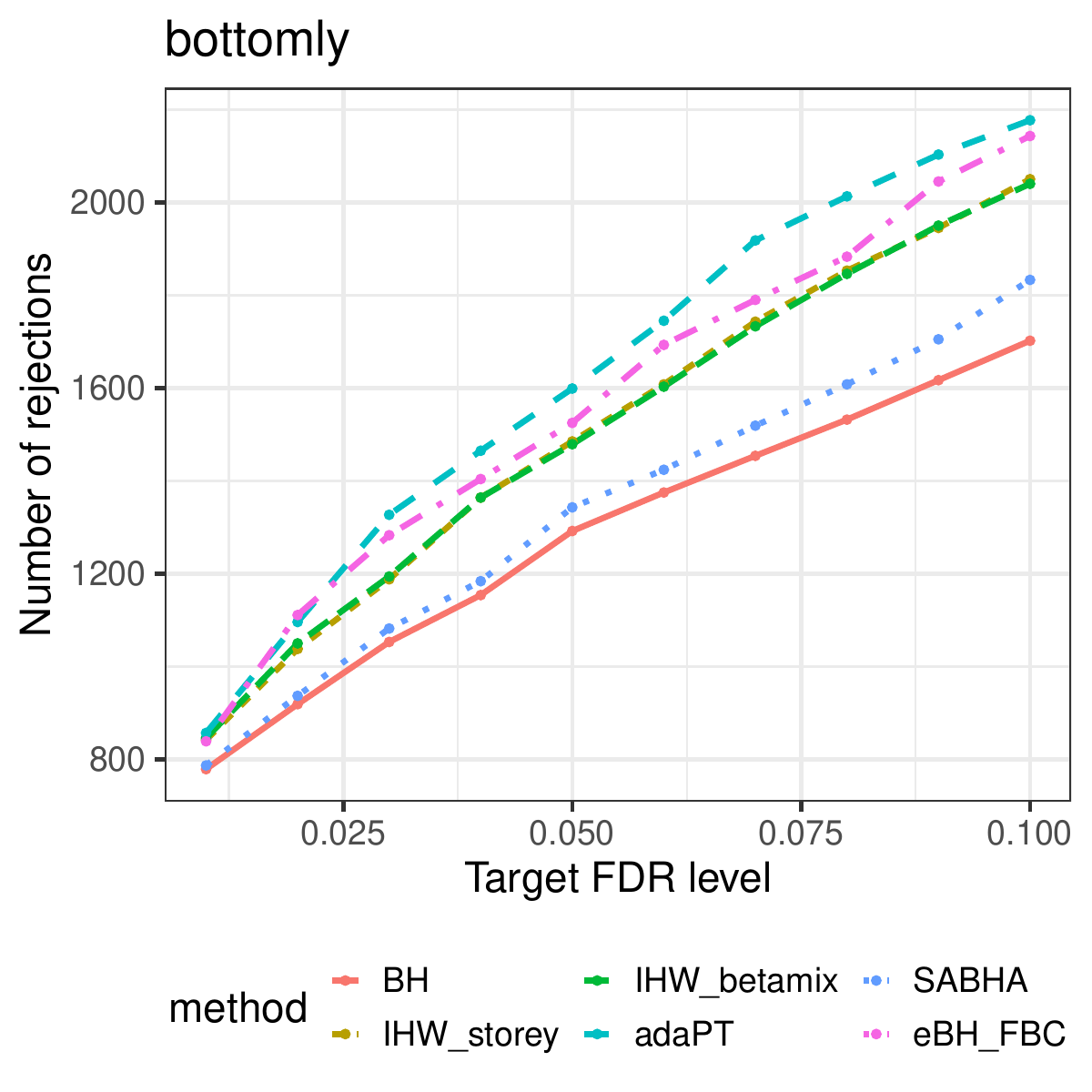}
    \includegraphics[width = 0.3\textwidth]{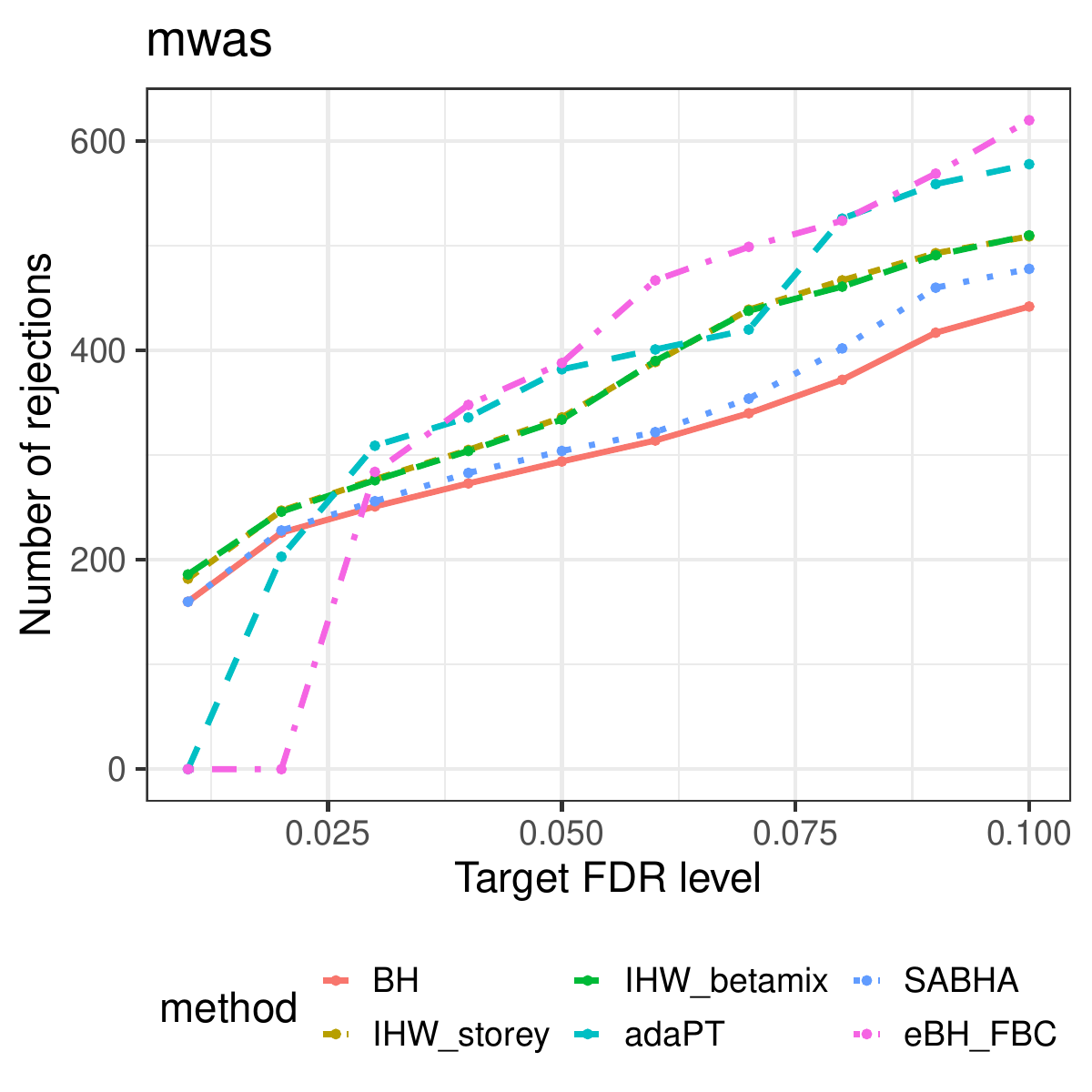}
    \caption{Number of discoveries of various methods with the target FDR level ranging from $0.01$ to $0.1$ in three real datasets. }
    \label{fig:ebh-gbc-real}
\end{figure}

The results of different methods for target FDR levels ranging from $0.01$ to $0.1$ are presented in Figure~\ref{fig:ebh-gbc-real}. \texttt{AdaPT} and \texttt{eBH\_FBC} are the two methods that make the most discoveries (except for the MWAS data set with a target FDR level below 0.025). %Both versions of \texttt{IHW} outperform \texttt{SABHA} and the BH procedure.
For the airway dataset, \texttt{AdaPT} consistently produces the most discoveries at higher target FDR levels, which is due to the high signal density of this dataset. For instance, when the target FDR level is 10\%, \texttt{AdaPT} is able to identify 6,053 discoveries out of 18,028 tests. It is worth noting that the proposed method performs similarly to \texttt{AdaPT} when the target FDR level is below 4\%. We observe a similar phenomenon for the Bottomly dataset. For the MWAS dataset, both \texttt{AdaPT} and \texttt{eBH\_FBC} fail to make any discoveries when the target FDR level is set to 1\%. This is a limitation of the BC-type method, which may have reduced power when the signal is very sparse. However, as the target FDR level increases, \texttt{AdaPT} and \texttt{eBH\_FBC} quickly outperform the other methods in terms of the number of discoveries. \texttt{eBH\_FBC} outperforms AdaPT with more discoveries when the FDR level is above 5\%. Overall, \texttt{eBH\_FBC} performs comparably to \texttt{AdaPT}.

\section{Discussions}\label{sec:con}
Motivated by the recent findings in \citet{li2025note}, we transform testing results from different procedures or data subsets into e-values. By aggregating these e-values, we obtain a combined set of e-values that captures information across procedures or partitions. A key feature of our method is the use of data-dependent weights, constructed via a leave-one-out approach, to ensure the resulting e-values yield finite-sample FDR control under the e-BH procedure. This weighted version is often more powerful than its unweighted counterpart. Simulations further reveal that a computationally efficient approximation of the weights achieves comparable performance.

We envision the idea of aggregating different multiple testing results through e-values to be useful in other contexts, such as meta-analysis or federated learning. Other interesting future research problems include finding the optimal way of combining the e-values with respect to certain criteria and investigating the robustness of the proposed methods when the data exhibit dependence.

\begin{center}
{\large\bf SUPPLEMENTARY MATERIAL}
\end{center}
\begin{description}
\item[Supplement:] Including all the proofs, additional discussions, and numerical results.
\end{description}

\bibliographystyle{plainnat}
\bibliography{reference.bib}

\newpage
\appendix
\begin{center}
\textbf{\LARGE Appendices}    
\end{center}
\renewcommand\thesection{\Alph{section}}
\numberwithin{equation}{section}
\numberwithin{table}{section}
\numberwithin{figure}{section}
\numberwithin{theorem}{section}
\numberwithin{lemma}{section}
\numberwithin{proposition}{section}

\section{Preliminaries}
\subsection{Unified Form of Multiple Testing Procedures}
Table~\ref{tab:uni-form} summarizes the specifications of $m(t)$ and $R_i(t)$ for different multiple testing procedures.
\begin{table}[h!]
    \centering
    \begin{tabular}{|c|cc|}
    \hline
    Method & $m(t)$ & $R_i(t)$ \\
    \hline
    BH & $n t$ & $\ind\{p_i \leq t\}$ \\
    ST & $n \pi_0^{\lambda} t$ & $\ind\{p_i \leq t\}$ \\
    BC & $1 + \sum_{i=1}^n \ind\{p_i \geq 1 - t\}$ & $\ind\{p_i \leq t\}$ \\
    FBH & $n g(t)$ & $\ind\{\varphi_i(p_i) \leq t\}$ \\
    FBC & $1 + \sum_{i=1}^n \ind\{\varphi_i(1 - p_i) \leq t\}$ & $\ind\{\varphi_i(p_i) \leq t\}$ \\
    \hline
    \end{tabular}
    \caption{Definitions of $m(t)$ and $R_i(t)$ for various multiple testing procedures.}
    \label{tab:uni-form}
\end{table}

\section{Proofs of Main Results}   
We begin by stating the following propositions, whose proofs are deferred to Appendix~\ref{appe:add_proof} to ensure self-containment. These results will be frequently utilized in the subsequent proofs of the main theorems.

\begin{proposition}[Lemma 6 of \citet{barber2020robust}]\label{prop:lemma-bcbh}
Let $T_{\BC,i}$ be the threshold for the BC methods when $ p_i $ is replaced with $\min\{p_i, 1- p_i\}$. For any $i$, $j$, if $\min(p_i,p_j) \geq 1 - \max\{T_{\BC,i}, T_{\BC,j}\}$, then we have $T_{\BC,i} = T_{\BC,j}$.
\end{proposition}

\begin{proposition}[Proposition A.2 of \citet{li2025note}]\label{prop:gbcthres}
Suppose that the null p-values are mutually independent, are independent of the alternative p-values, and satisfy Condition~\eqref{eq:ass-p-bc}. Let $T_i$ be the threshold for the generalized BC methods when $ p_i $ is replaced with $\min\{p_i, 1- p_i\}$. For any $i$, $j$, if $\max\{\varphi_i(1 - p_i), \varphi_j(1 - p_j)\} \leq \max\{T_i, T_j\}$, then we have $T_i = T_j$.
\end{proposition}

\subsection{Proof of Theorem~\ref{thm:bcbh-weight}}\label{proof:thm:bcbh-weight}
\begin{proof}
Consider the BH procedure and observe that for a given number of rejections $R_{\BH}$, $T_{\BH} = T_{\BH}(R_{\BH})$ is a deterministic function of $R_{\BH}$. Let $R_{\BH}(p_i\rightarrow 0)$ be the number of rejections obtained by replacing the p-value $p_i$ with 0.
Using the above fact and the leave-one-out argument, we have
\begin{align*}
&\sum_{i\in\cH_0}\bbE[w_{\BH, i}e_{\BH, i}]\\
=& \sum_{i\in\cH_0}\bbE\left[\frac{T_{\BH, i}}{T_{\BH, i} + \frac{1}{n}\left(1 + \sum_{j\neq i}\ind\{p_j \geq 1 - T_{\BC, j,i}\} \right)}\frac{1}{T_{\BH}}\ind\{p_i \leq T_{\BH}\}\right]\\
=& \sum_{i\in\cH_0}\sum_{k=1}^n\bbE\left[\frac{T_{\BH, i}}{T_{\BH, i} + \frac{1}{n}\left(1 + \sum_{j\neq i}\ind\{p_j \geq 1 - T_{\BC, j,i}\} \right)}\frac{1}{T_{\BH}(k)}\ind\{p_i \leq T_{\BH}(k), R_{\BH}(p_i\rightarrow 0) = k\}\right],
\end{align*}
where to get the second equality, we have used the fact that when the $i$th hypothesis is rejected (i.e., $p_i\leq T_{\BH}$), $R_{\BH}=R_{\BH}(p_i\rightarrow 0)$.
Let $\cF_i$ be the sigma algebra generated by $\{p_1, \cdots, p_{i-1}, 0, p_{i+1}, \cdots, p_n\}$. We have
\begin{align*}
&\sum_{k = 1}^n \bbE\left[\frac{T_{\BH, i}}{T_{\BH, i} + \frac{1}{n}\left(1 + \sum_{j\neq i}\ind\{p_j \geq 1 - T_{\BC, j,i}\} \right)}\frac{1}{T_{\BH}(k)}\ind\{p_i \leq T_{\BH}(k), R_{\BH}(p_i\rightarrow 0) = k\} \Bigg| \cF_i\right]\\
=& \sum_{k=1}^n \frac{T_{\BH, i}}{T_{\BH, i} + \frac{1}{n}\left(1 + \sum_{j\neq i}\ind\{p_j \geq 1 - T_{\BC, j,i}\} \right)}\frac{1}{T_{\BH}(k)} P(p_i \leq T_{\BH}(k)) \bbE[\ind\{R_{\BH}(p_i\rightarrow 0) = k\}\lvert \cF_i]\\
\leq&\frac{T_{\BH, i}}{T_{\BH, i} + \frac{1}{n}\left(1 + \sum_{j\neq i}\ind\{p_j \geq 1 - T_{\BC, j,i}\} \right)},
\end{align*}
where we used the fact that $T_{\BH, i}$ and $T_{\BC, j, i}$ are both measurable with respect to $\mathcal{F}_i$. Thus,
\begin{align*}
&\sum_{i\in\cH_0}\bbE[w_{\BH, i}e_{\BH, i}]\\
\leq&\sum_{i\in\cH_0}\bbE\left[\frac{T_{\BH, i}}{T_{\BH, i} + \frac{1}{n}\left(1 + \sum_{j\neq i}\ind\{p_j \geq 1 - T_{\BC, j,i}\} \right)}\right]\\
\leq &\sum_{i\in\cH_0}\bbE\left[\frac{\max_iT_{\BH, i}}{\max_iT_{\BH, i} + \frac{1}{n}\left(1 + \sum_{j\neq i}\ind\{p_j \geq 1 - T_{\BC, j, i}\} \right)}\right].
\end{align*}
Note that $T_{\BC,j,i}\geq T_{\BC,j}$ and hence $\ind\{p_j \geq 1 - T_{\BC,j, i}\}\geq \ind\{p_j \geq 1 - T_{\BC, j}\}$. It follows that
\begin{align*}
\sum_{i\in\cH_0}\bbE[w_{\BH, i}e_{\BH, i}]
\leq &\sum_{i\in\cH_0}\bbE\left[\frac{\max_iT_{\BH, i}}{\max_iT_{\BH, i} + \frac{1}{n}\left(1 + \sum_{j\neq i}\ind\{p_j \geq 1 - T_{\BC, j}\} \right)}\right]
\\ \leq &\bbE\left[\frac{ n \max_iT_{\BH, i}}{\max_iT_{\BH, i} + \frac{1}{n}\sum_{j=1}^n\ind\{p_j \geq 1 - T_{\BC, j}\}}\right].
\end{align*}
For the BC procedure, let $\widetilde \cF_i$ be the sigma algebra generated by $\bfp_{-i}$. Then, we have
\begin{align*}
&\sum_{i\in\cH_0}\bbE[w_{\BC, i}e_{\BC, i}]\\
=& \sum_{i\in\cH_0}\bbE\left[\frac{\frac{1}{n}\left(1 + \sum_{j\neq i}\ind\{p_j \geq 1 - T_{\BC}\}\right)}{\max_iT_{\BH, i} + \frac{1}{n}\left(1 + \sum_{j\neq i}\ind\{p_j \geq 1 - T_{\BC}\} \right)} \frac{n\ind\{p_i \leq T_{\BC}\}}{1 + \sum_{j=1}^n\ind\{p_j \geq 1 - T_{\BC}\}}\right]\\
=& \sum_{i\in\cH_0}\bbE\left[\frac{\frac{1}{n}\left(1 + \sum_{j\neq i}\ind\{p_j \geq 1 - T_{\BC, i}\}\right)}{\max_iT_{\BH, i} + \frac{1}{n}\left(1 + \sum_{j\neq i}\ind\{p_j \geq 1 - T_{\BC, i}\} \right)} \frac{n\ind\{p_i \leq T_{\BC, i}\}}{1 + \sum_{j\neq i}\ind\{p_j \geq 1 - T_{\BC, i}\}}\right]\\
=& \sum_{i\in\cH_0}\bbE\left[\frac{1}{\max_iT_{\BH, i} + \frac{1}{n}\left(1 + \sum_{j\neq i}\ind\{p_j \geq 1 - T_{\BC, i}\} \right)} \bbE[\ind\{p_i \leq T_{\BC, i}\}\lvert \widetilde \cF_i] \right]\\
\leq& \sum_{i\in\cH_0}\bbE\left[\frac{\ind\{p_i \geq 1 - T_{\BC, i}\}}{\max_iT_{\BH, i} + \frac{1}{n}\left(1 + \sum_{j\neq i}\ind\{p_j \geq 1 - T_{\BC, i}\} \right)}\right],
\end{align*}
where (i) we have used the fact that $T_{\BC} = T_{\BC, i}$ when $p_i \leq T_{BC} < 0.5$ to get the second equation, (ii) the third equation follows because both $\max_iT_{\BH, i}$ and $T_{\BC,i}$ are measurable with respect to $\widetilde \cF_i$, and (iii) the inequality is due to the assumption that $p_i$ follows the super-uniform distribution on $[0,1]$ and thus satisfies Condition \eqref{eq:ass-p-bc}.

By Proposition~\ref{prop:lemma-bcbh}, we have
\begin{equation}\label{eq:change-index}
\frac{\ind\{p_i \geq 1 - T_{\BC, i}\}}{\max_iT_{\BH, i} + \frac{1}{n}\left(1 + \sum_{j\neq i}\ind\{p_j \geq 1 - T_{\BC, i}\} \right)} = \frac{ \ind\{ p_i \geq 1 - T_{\BC, i}\}}{\max_iT_{\BH, i} + \frac{1}{n}\left( \sum_{j=1}^n\ind\{p_j \geq 1 - T_{\BC, j}\} \right)}.    
\end{equation}
If $p_i < 1 - T_{\BC, i}$, both sides are equal to $0$. If $p_i \geq 1 - T_{\BC, i}$, we claim that $\ind\{p_j \geq 1 - T_{\BC, i}\} = \ind\{p_j \geq 1 - T_{\BC, j}\}$. Indeed, if $p_j \geq 1 - T_{\BC, i}$ but $p_j < 1 - T_{\BC, j}$, we have $T_{\BC, i} > T_{\BC,j}$. This implies that $\min(p_i,p_j) \geq 1 - \max\{T_{\BC,i}, T_{\BC,j}\}$. By Proposition~\ref{prop:lemma-bcbh}, we have $T_{\BC, i} = T_{\BC, j}$, which contradicts with the fact that $T_{\BC, i} > T_{\BC, j}$. The other direction can be proved similarly. Hence, we have
\begin{align*}
\sum_{i\in\cH_0}\bbE[w_{\BC, i}e_{\BC, i}] =& \bbE\left[\frac{\sum_{i\in\cH_0}\ind\{p_i \geq 1 - T_{\BC, i}\}}{\max_iT_{\BH, i} + \frac{1}{n}\left(1 + \sum_{j\neq i}\ind\{p_j \geq 1 - T_{\BC, j}\} \right)}\right]
\\ \leq & \bbE\left[\frac{\sum_{i=1}^n \ind\{p_i \geq 1 - T_{\BC, i}\}}{\max_iT_{\BH, i} + \frac{1}{n}\sum_{j=1 }^n\ind\{p_j \geq 1 - T_{\BC, j}\}}\right].
\end{align*}
Combining the arguments, we obtain
\begin{align*}
&\sum_{i\in\cH_0}\{\bbE[w_{\BH, i}e_{\BH, i}]+\bbE[w_{\BC, i}e_{\BC, i}]\}
\leq \bbE\left[\frac{n\max_iT_{\BH, i}+\sum_{i=1}^n \ind\{p_i \geq 1 - T_{\BC, i}\}}{\max_iT_{\BH, i} + \frac{1}{n}\sum_{j=1 }^n\ind\{p_j \geq 1 - T_{\BC, j}\}}\right]= n.
\end{align*}
\end{proof}

\subsection{Proof of Theorem~\ref{thm:fair-weight}}\label{proof:fair-weight}
\begin{proof}
We only present the proof for the case of $G = 2$. The arguments can be generalized to the general case without essential difficulty. Let us consider the first group.
 Let $\cF_i$ be the sigma algebra generated by $\bfp_{1, i}$. Then, we have
\begin{align*}
&\sum_{i\in\cG_1\cap\cH_0}\bbE[w_{i}e_{i}]\\
=&\sum_{i\in\cG_1\cap\cH_0}\bbE\left[\frac{\frac{n}{n_1}\left(1 + \sum_{j\neq i, j\in\cG_1}\ind\{p_{j}\geq 1 - T_1\}\right)}{\left(1 + \sum_{j\neq i, j\in\cG_1}\ind\{p_{j}\geq 1 - T_1\}\right) + \sum_{j\in\cG_2}\ind\{p_{j} \geq 1 - T_{2, j}\}}\frac{n_1\ind\{p_{i}\leq T_1\}}{1 + \sum_{j\in\cG_1}\ind\{p_{j}\geq 1 - T_1\}}\right]\\
=&\sum_{i\in\cG_1\cap\cH_0}\bbE\left[\frac{n\ind\{p_{i}\leq T_{1, i}\}}{\left(1 + \sum_{j\neq i, j\in\cG_1}\ind\{p_{j}\geq 1 - T_{1, i}\}\right) + \sum_{j\in\cG_2}\ind\{p_{j} \geq 1 - T_{2, j}\}}\right]\\
=&\sum_{i\in\cG_1\cap\cH_0}\bbE\left[\frac{n}{\left(1 + \sum_{j\neq i, j\in\cG_1}\ind\{p_{j}\geq 1 - T_{1, i}\}\right) + \sum_{j\in\cG_2}\ind\{p_{j} \geq 1 - T_{2, j}\}}\bbE[\ind\{p_{i}\leq T_{1, i}\} | \cF_i]\right]\\
\leq&\sum_{i\in\cG_1\cap\cH_0}\bbE\left[\frac{n\ind\{p_{i}\geq 1 - T_{1, i}\}}{\left(1 + \sum_{j\neq i, j\in\cG_1}\ind\{p_{j}\geq 1 - T_{1, i}\}\right) + \sum_{j\in\cG_2}\ind\{p_{j} \geq 1 - T_{2, j}\}}\right],
\end{align*}
where (i) we have used the fact that $T_1 = T_{1, i}$ when $p_i \leq T_1 < 0.5$ to get the second equation, (ii) the third equation follows because $T_{1, i}$ are measurable with respect to $\cF_i$, and (iii) the inequality is due to the assumption that $p_i$ satisfies Condition~\eqref{eq:ass-p-bc} under the null.

By Proposition~\ref{prop:lemma-bcbh} and the proof of \eqref{eq:change-index}, we have
\begin{align*}
&\frac{n\ind\{p_{i}\geq 1 - T_{1, i}\}}{\left(1 + \sum_{j\neq i, j\in\cG_1}\ind\{p_{j}\geq 1 - T_{1, i}\}\right) + \sum_{j\in\cG_2}\ind\{p_{j} \geq 1 - T_{2, j}\}} \\
=& \frac{n\ind\{p_{i}\geq 1 - T_{1, i}\}}{\sum_{j\in\cG_1}\ind\{p_{j}\geq 1 - T_{1, j}\} + \sum_{j\in\cG_2}\ind\{p_{j} \geq 1 - T_{2, j}\}}.
\end{align*}
Thus,
\[\sum_{i\in\cH_0\cap\cG_1}\bbE[w_{i}e_{i}] \leq \frac{n\sum_{i\in\cG_1}\ind\{p_{i}\geq 1 - T_{1, i}\}}{\sum_{j\in\cG_1}\ind\{p_{j}\geq 1 - T_{1, j}\} + \sum_{j\in\cG_2}\ind\{p_{2, j} \geq 1 - T_{2, j}\}}.\]
Using the same argument for the second group, we obtain
\[\sum_{i\in\cH_0\cap\cG_2}\bbE[w_{i}e_{i}] \leq \frac{n\sum_{i\in\cG_2}\ind\{p_{i}\geq 1 - T_{2, i}\}}{\sum_{j\in\cG_1}\ind\{p_{j}\geq 1 - T_{1, j}\} + \sum_{j\in\cG_2}\ind\{p_{2, j} \geq 1 - T_{2, j}\}}.\]
Hence,
\[\sum_{i\in\cG_1\cap\cH_0}\bbE[w_{i}e_{i}] + \sum_{i\in\cG_2\cap\cH_0}\bbE[w_{i}e_{i}] \leq n.\]
\end{proof}

\subsection{Proof of Theorem~\ref{thm:struc-nodata}}\label{proof:struc-nodata}
\begin{proof}
We only prove the result for $G = 2$ (the same argument applies to the case of a general $G$). Note that when $\hat \varphi_i(p_i)\leq T_g \leq T_{g, \up} < \hat \varphi_i(0.5)$, Assumption 2 implies that $p_i < 0.5$ and hence $p_i = \tilde p_i$. By the definition of $T_{g, i}$, $T_g = T_{g, i}$. Therefore, for the first group, we have
\begin{align*}
\sum_{i\in\cH_0\cap\cG_1}\bbE[w_i e_i] =& \sum_{i\in\cH_0\cap\cG_1} \bbE\left[\frac{n_1 w_i\ind\{\hat \varphi_i(p_i)\leq T_1\}}{1 + \sum_{j\in\cG_1}\ind\{\hat \varphi_j(1 - p_i)\leq T_1\}}\right]\\
=& \sum_{i\in\cH_0\cap\cG_1} \bbE\left[\frac{n_1 w_i \ind\{\hat \varphi_i(p_i)\leq T_{1, i}\}}{1 + \sum_{j\in\cG_1, j \neq i}\ind\{\hat \varphi_j(1 - p_i)\leq T_{1, i}\}}\right].
\end{align*}
Let $\cF_{i}$ denote the sigma algebra generated by $\bfp_{-i}$. Since $T_{1, i}$, $\hat \varphi_{j}$ and $w_i$ are all measurable with respect to $\cF_i$, we deduce that
\begin{align*}
\sum_{i\in\cH_0\cap\cG_1}\bbE[w_i e_i] =& \sum_{i\in\cH_0\cap\cG_1} \bbE\left[\frac{n_1 w_i}{1 + \sum_{j\in\cG_1, j\neq i}\ind\{\hat \varphi_j(1 - p_i)\leq T_{1, i}\}}\bbE[\ind\{\hat \varphi_i(p_i)\leq T_{1, i}\}\lvert \cF_i]\right]\\
\leq & \sum_{i\in\cH_0\cap\cG_1} \bbE\left[\frac{n_1 w_i}{1 + \sum_{j\in\cG_1, j\neq i}\ind\{\hat \varphi_j(1 - p_i)\leq T_{1, i}\}}\bbE[\ind\{\hat \varphi_i(1 - p_i)\leq T_{1, i}\}\lvert \cF_i]\right]\\
=& \sum_{i\in\cH_0\cap\cG_1} \bbE\left[\frac{n_1 w_i\ind\{\hat \varphi_i(1 - p_i)\leq T_{1, i}\}}{1 + \sum_{j\in\cG_1, j\neq i}\ind\{\hat \varphi_j(1 - p_i)\leq T_{1, i}\}}\right],
\end{align*}
where we use Assumption 1(C) to get the inequality.

By Proposition ~\ref{prop:gbcthres} and Assumption 2, we have
\begin{align*}
\frac{\ind\{\hat \varphi_i(1 - p_i)\leq T_{1, i}\}}{1 + \sum_{j\in\cG_1, j\neq i}\ind\{\hat \varphi_j(1 - p_j)\leq T_{1, i}\}} &= \frac{\ind\{\hat \varphi_i(1 - p_i)\leq T_{1, i}\}}{1 + \sum_{j\in\cG_1, j\neq i}\ind\{\hat \varphi_j(1 - p_j)\leq T_{1, j}\}}\\
&= \frac{\ind\{\hat \varphi_i(1 - p_i)\leq T_{1, i}\}}{\sum_{j\in\cG_1} \ind\{\hat \varphi_j(1 - p_j)\leq T_{1, j}\}}.    
\end{align*}
If $\hat \varphi_i(1 - p_i) > T_{1, i}$, both sides are equal to $0$. If $\hat \varphi_i(1 - p_i) \leq T_{1, i}$, we claim that $\ind\{\hat \varphi_j(1 - p_j) \leq T_{1, i}\} = \ind\{\hat \varphi_j(1 - p_j) \leq T_{1, j}\}$. Indeed, if $\hat \varphi_j(1 - p_j) > T_{1, i}$ but $\hat \varphi_j(1 - p_j) \leq T_{1, j}$, then we have $T_{1, i} < T_{1, j}$. Hence, $\hat \varphi_i(1 - p_i) \leq T_{1, i} < T_{1, j}$. By proposition~\ref{prop:gbcthres}, we have $T_{1, i} = T_{1, j}$, which contradicts with $T_{1, i} < T_{1, j}$.
The other direction can be proved similarly. 

If the weights $\{w_i\}$ are independent of the p-values and covariate information, then we have
\[\sum_{i\in\cG_1\cap\cH_0}\bbE[w_i e_i] \leq n_1\max_{i\in\cG_1}w_i \bbE\left[\frac{\sum_{i\in\cG_1\cap\cH_0}\ind\{\hat \varphi_i(1 - p_i)\leq T_{1, i}\}}{\sum_{j\in\cG_1}\ind\{\hat \varphi_j(1 - p_j)\leq T_{1, j}\}}\right] \leq n_1\max_{i\in\cG_1}w_i.\]
Using the same argument for the second group, we obtain
\[\sum_{i\in\cG_2\cap\cH_0}\bbE[w_i e_i]\leq n_2\max_{i\in\cG_2}w_i.\]
Hence, by \eqref{eq:cond-weight-cross}, we deduce that
\[\sum_{i\in\cH_0}\bbE[w_i] \leq n.\]
\end{proof}

\subsection{Proof of Theorem~\ref{thm:struc-evalue}}\label{proof:struc-evalue}
\begin{proof}
We only prove the result for $G = 2$ (the same argument applies to the case of a general $G$). Note that when $\hat \varphi_i(p_i)\leq T_g \leq T_{g, \up} < \hat \varphi_i(0.5)$, Assumption 2 implies that $p_i < 0.5$ and thus $p_i = \tilde p_i$. Thus, $T_g = T_{g, i}$ by the definition of $T_{g, i}$. Therefore, for the first group, we have
\begin{align*}
&\sum_{i\in\cH_0\cap\cG_1}\bbE[w_{i}e_{i}]\\
=&
\sum_{i\in\cH_0\cap\cG_1}\bbE\left[\frac{\frac{n}{n_1}\left(1 + \sum_{j\neq i, j\in\cG_1}\ind\{\hat \varphi_{j}(1 - p_{j})\leq T_1\}\right)}{\left(1 + \sum_{j\neq i, j\in\cG_1}\ind\{\hat \varphi_{j}(1 - p_{j})\leq T_1\}\right) + \sup_{p\in[0, 1]}\sum_{j\in\cG_2}\ind\{\hat \varphi_{j}^{1, i, p}(1 - p_{j}) \leq T_{2, j}^{1, i, p}\}}\right.\ \\
&\qquad\times\left.\frac{n_1\ind\{\hat \varphi_{i}(p_{i})\leq T_1\}}{1 + \sum_{j\in\cG_1}\ind\{\hat \varphi_{j}(1 - p_{j}) \leq T_1\}}\right]\\
\leq &\sum_{i\in\cH_0\cap\cG_1}\bbE\left[\frac{n\ind\{\hat \varphi_{i}(p_{i}) \leq T_{1, i}\}}{\left(1 + \sum_{j\neq i, j\in\cG_1}\ind\{\hat \varphi_{j}(1 - p_{j})\leq T_{1, i}\}\right) + \sup_{p\in[0, 1]}\sum_{j\in\cG_2}\ind\{\hat \varphi_{j}^{1, i, p}(1 - p_{j}) \leq T_{2, j}^{1, i, p}\}}\right].
\end{align*}
Let $\cF_{i}$ denote the sigma algebra generated by $\bfp_{-i}$. Since $T_{1, i}$, $\hat \varphi_{j}$ for $j\in\cG_1$, $\hat \varphi_{j}^{1, i, p}$, and $T_{2, j}^{1, i, p}$ for $j\in\cG_2$ are all measurable with respect to $\cF_i$, we deduce that
\begin{align*}
&\sum_{i\in\cH_0\cap\cG_1}\bbE[w_{i}e_{i}]\\
=&\sum_{i\in\cH_0\cap\cG_1}\bbE\left[\bbE\left[\frac{n\ind\{\hat \varphi_{i}(p_{i}) \leq T_{1, i}\}}{\left(1 + \sum_{j\neq i, j\in\cG_1}\ind\{\hat \varphi_{j}(1 - p_{j})\leq T_{1, i}\}\right) + \sup_{p\in[0, 1]}\sum_{j\in\cG_2}\ind\{\hat \varphi_{j}^{1, i, p}(1 - p_{j}) \leq T_{2, j}^{1, i, p} \}}\Bigg\lvert\cF_{i}\right]\right]\\
\leq&\sum_{i\in\cH_0\cap\cG_1}\bbE\left[\bbE\left[\frac{n\ind\{\hat \varphi_{i}(1 - p_{i}) \leq T_{1, i}\}}{\left(1 + \sum_{j\neq i, j\in\cG_1}\ind\{\hat \varphi_{j}(1 - p_{j})\leq T_{1, i}\}\right) + \sup_{p\in[0, 1]}\sum_{j\in\cG_2}\ind\{\hat \varphi_{j}^{1, i, p}(1 - p_{j}) \leq T_{2, j}^{1, i, p}\}}\Bigg\lvert\cF_{i}\right]\right]\\
\leq&\sum_{i\in\cH_0\cap\cG_1}\bbE\left[\frac{n\ind\{\hat \varphi_{i}(1 - p_{i}) \leq T_{1, i}\}}{\left(1 + \sum_{j\neq i, j\in\cG_1}\ind\{\hat \varphi_{j}(1 - p_{j})\leq T_{1, i}\}\right) + \sum_{j\in\cG_2}\ind\{\hat \varphi_{j}(1 - p_{j}) \leq T_{2, j}\}}\right],
\end{align*}
where we have used Assumption 1(C) to obtain the first inequality, and the second inequality is due to the fact that
\begin{align*}
\sum_{j\in G_2}\ind\{\hat \varphi_j(1 - p_j)\leq T_{2, j}\} &= \sum_{j\in G_2}\ind\{\hat \varphi_j^{1, i, p}(1 - p_j)\leq T_{2, j}^{1, i, p}\}\lvert_{p = p_i}\\
&\leq \sup_{p\in[0, 1]}\sum_{j\in G_2}\ind\{\hat \varphi_j^{1, i, p}(1 - p_j)\leq T_{2, j}^{1, i, p}\}.
\end{align*}

By Proposition~\ref{prop:gbcthres} and the argument in the proof of Theorem~\ref{thm:struc-nodata}, we have
\begin{align*}
&\frac{n\ind\{\hat \varphi_{i}(1 - p_{i}) \leq T_{1, i}\}}{\left(1 + \sum_{j\neq i, j\in\cG_1}\ind\{\hat \varphi_{j}(1 - p_{j})\leq T_{1, i}\}\right) + \sum_{j\in\cG_2}\ind\{\hat \varphi_{j}(1 - p_{j}) \leq T_{2, j}\}}\\
=&\frac{n\ind\{\hat \varphi_{i}(1 - p_{i}) \leq T_{1, i}\}}{ \sum_{ j\in\cG_1}\ind\{\hat \varphi_{j}(1 - p_{j})\leq T_{1, j}\} + \sum_{j\in\cG_2}\ind\{\hat \varphi_{j}(1 - p_{j}) \leq T_{2, j}\}}.
\end{align*}
Hence, for the first group, we get
\begin{align*}
\sum_{i\in\cH_0\cap\cG_1}\bbE[w_{i}e_{i}] \leq \bbE\left[\frac{n\sum_{i\in\cG_1}\ind\{\hat \varphi_{i}(1 - p_{i}) \leq T_{1, i}\}}{\sum_{ j\in\cG_1}\ind\{\hat \varphi_{j}(1 - p_{j})\leq T_{1, j}\} + \sum_{j\in\cG_2}\ind\{\hat \varphi_{j}(1 - p_{j}) \leq T_{2, j}\}}\right].
\end{align*}
Following the same discussion, we have
\begin{align*}
\sum_{i\in\cH_0\cap\cG_2}\bbE[w_{i}e_{i}] \leq \bbE\left[\frac{n\sum_{i\in\cG_2}\ind\{\hat \varphi_{i}(1 - p_{i}) \leq T_{1, i}\}}{\sum_{ j\in\cG_1}\ind\{\hat \varphi_{j}(1 - p_{j})\leq T_{1, j}\} + \sum_{j\in\cG_2}\ind\{\hat \varphi_{j}(1 - p_{j}) \leq T_{2, j}\}}\right].
\end{align*}
Combining the above results leads to
\[\sum_{g=1}\sum_{i\in\cH_0\cap\cG_g}\bbE[w_{i}e_{i}] \leq n.\]
\end{proof}

\section{Supplementary Proofs}\label{appe:add_proof}
\subsection{Proof of Proposition~\ref{prop:ebh-control}}\label{proof:ebh-control}
\begin{proof}
Note that
\begin{align*}
\text{FDP}=&\sum_{i=1}^n\frac{\ind\{ie_{(i)}\geq n/\alpha,H_{(i)} \text{ is under the null}\}}{1\vee \hat{k}} 
\\ \leq&\sum_{i=1}^n\frac{\ind\{ie_{(i)}\geq n/\alpha,H_{(i)} \text{ is under the null}\}}{1\vee i} 
\\ \leq&\sum_{i=1}^n\ind\{H_{(i)} \text{ is under the null}\}\frac{\alpha e_{(i)}}{n}=\frac{\alpha}{n}\sum_{i\in\mathcal{H}_0}e_i.
\end{align*}
Under Condition~\eqref{eq:evalue}, we have
\begin{align*}
\text{FDR}=\mathbb{E}[\text{FDP}]\leq \alpha.  
\end{align*}
\end{proof}

\subsection{Proof of Proposition~\ref{prop:lemma-bcbh}}
\begin{proof}
Proposition \ref{prop:lemma-bcbh} is a special case of Proposition~\ref{prop:gbcthres} by choosing $\varphi_i$ as the identity function for all $1 \leq i \leq n$.
\end{proof}

\subsection{Proof of Proposition~\ref{prop:gbcthres}}
\begin{proof}
Write $T = T_{\fbc}$ for the ease of notation. First, given a p-value vector $\bfp = (p_1, \cdots, p_n)$, recall that the threshold $T$ is defined as
\[T = \max\left\{0 < t \leq T_{\up}\colon \underbrace{\frac{1 + \sum_{l=1}^n\ind\{\varphi_l(1 - p_l) \leq t\}}{\sum_{l=1}^n \ind\{\varphi_l(p_l)\leq t\}}}_{\coloneqq g(\bfp, t)}\leq \alpha \right\},\]
where $T_{\up}$ satisfies $T_{\up} < \varphi_l(0.5)$ for all $l$.

Without loss of generality, let us assume $T_i \geq T_j$. By the assumption that $\max\{\varphi_i(1 - p_i), \varphi_j(1 - p_j)\} \leq \max\{T_i, T_j\}$, we have $\varphi_i(1 - p_i) \leq T_i$ and $\varphi_j(1 - p_j) \leq T_i$. Since $\varphi_i$ is an increasing function, we have $\varphi_i(1 - p_i) \leq T_{\up} < \varphi_i(0.5)$, which implies $1 - p_i < 0.5$. Thus $\varphi_i(p_i) \geq \varphi_i(0.5) > T_{\up} \geq T_i$. The same discussion for $p_j$ leads to $\varphi_j(p_j) > T_i$. 

Denote $\tilde p_i = \min\{p_i, 1 - p_i\}$ and $\bfp_{-i} = (p_1, \cdots, p_{i-1}, \tilde p_i, p_{i+1}, \cdots, p_n)$ for all $i$. Consider the function
\[g(\bfp_{-j}, T_i) = \frac{1 + \sum_{l=1}^n\ind\{\varphi_l(1 - p_{-j, l})\leq T_i\}}{\sum_{l=1}^n\ind\{\varphi_l(p_{-j, l})\leq T_i\}},\]
where $p_{-j, l}$ is the $l$th entry of $p_{-j}$.
For the denominator, we have
\begin{align*}
&\sum_{l = 1}^n \ind\{\varphi_l(p_{-j, l})\leq T_i\}\\
=& \sum_{l = 1}^n \ind\{\varphi_l(p_{-i, l}) \leq T_i\} + \underbrace{\ind\{\varphi_j(p_{-j, j})\leq T_i\}}_{=1} + \underbrace{\ind\{\varphi_i(p_{-j, i})\leq T_i\}}_{=0} \\
&\qquad - \underbrace{\ind\{\varphi_j(p_{-i, j})\leq T_i\}}_{=0} - \underbrace{\ind\{\varphi_i(p_{-i, i})\leq T_i\}}_{=1}\\
=&\sum_{l = 1}^n \ind\{\varphi_l(p_{-i, l}) \leq T_i\}.
\end{align*}
Similarly, for the numerator, we have
\begin{align*}
&\sum_{l = 1}^n \ind\{\varphi_l(1 - p_{-j, l})\leq T_i\}\\
=& \sum_{l = 1}^n \ind\{\varphi_l(1 - p_{-i, l}) \leq  T_i\} + \underbrace{\ind\{\varphi_j(1 - p_{-j, j}) \leq T_i\}}_{=0} \\
&\qquad + \underbrace{\ind\{\varphi_i(1 - p_{-j, i}) \leq  T_i\}}_{=1} - \underbrace{\ind\{\varphi_j(1 - p_{-i, j}) \leq  T_i\}}_{=1} - \underbrace{\ind\{\varphi_i(1 - p_{-i, i}) \leq T_i\}}_{0}\\
=&\sum_{l = 1}^n \ind\{\varphi_l(1 - p_{-i, l}) \leq T_i\}.
\end{align*}
Hence, $g(\bfp_{-j}, T_i) = g(\bfp_{-i}, T_i) \leq \alpha$. By the definition of $T_j$, we must have $T_i \leq T_j$. Similarly, we get $T_j\leq T_i$ and hence $T_i = T_j$.
\end{proof}

\newpage
\section{Assembling E-Values from Data Subsets}

We first compare our definition, which simultaneously controls group-wise and overall FDR, with the notion of predictive parity from the classification context in the fairness literature \citep{chouldechova2017fair}. To elaborate, consider a binary classification problem where $Y\in{0, 1}$ represents the true labels and $\widehat{Y}\in{0, 1}$ denotes the predicted labels. In this setting, the FDR is defined as $P(Y = 0 | \widehat{Y} = 1)$, and predictive parity requires equal FDR across all groups. However, the multiple testing scenario differs fundamentally from the classification problem, as the underlying truth of each hypothesis is unobserved and thus cannot directly inform the decision rule. Hence, in our context, it is more appropriate to control the FDRs across different groups at a common target level rather than enforcing strict equality.

\subsection{Additional Numerical Results: Group-Wise and Overall FDR Control}\label{appe:fair-examp}
\subsubsection*{Example: Data-Dependent Weights}
% {\color{red} Maybe add one or two sentences here. one or two sentences here.}
We begin with a concrete example to illustrate the form of the data-dependent weights. When $L = 2$, we have
\[w_i = \frac{\frac{n}{n_1}\left(1 + \sum_{j\neq i, j\in\cG_1}\ind\{p_{j}\geq 1 - T_1\}\right)}{\left(1 + \sum_{j\neq i, j\in\cG_1}\ind\{p_{j} \geq 1 - T_1\}\right) + \sum_{j\in\cG_2}\ind\{p_{j} \geq 1 - T_{2, j}\}}\]
for $i\in\cG_1$ and 
\[w_{i} = \frac{\frac{n}{n_2}\left(1 + \sum_{j\neq i, j\in\cG_2}\ind\{p_{j}\geq 1 - T_2\}\right)}{\sum_{j\in\cG_1}\ind\{p_{j}\geq 1 - T_{1, j}\}+\left(1 + \sum_{j\neq i, j\in\cG_2}\ind\{p_{j}\geq 1 - T_2\}\right)}\]
for $i\in\cG_2$.

\subsubsection*{A Naive Method that Controls both Group-wise and Overall FDR}
We tried BC\_Sep at the level $\alpha/G$, a naive method that controls both the group-wise and overall FDRs. Indeed, let $\hat n_{la}$ be the number of rejections for the $l$th group, and denote the number of false rejections for the $l$th group by $\hat n_{l0}$. Then we have $\bbE[\hat n_{l0} / (1\vee \hat n_{la})] \leq \alpha / L$, which implies that 
\[\bbE\left[\frac{\sum_{l=1}^L \hat n_{l0}}{1\vee \sum_{l=1}^L \hat n_{la}}\right] \leq \bbE\left[\sum_{l=1}^L \frac{\hat n_{l0}}{1\vee \hat n_{la}}\right] \leq \alpha.\]
However, this method has nearly zero power in all our simulation settings. Therefore, we have decided not to include its results in Table~\ref{tab:fair-exp}.

\subsubsection*{Parameter Values for the Two-Group Setting}
The parameter values for different settings for two groups are detailed in Table~\ref{tab:fair-para}.
\begin{table}
    \centering
    \begin{tabular}{|c|cccc|cccc|}
    \hline
     Setting & $n_1$ & $n_{1a}$ & $\alpha_1$ & $\beta_1$ & $n_2$ & $n_{2a}$ & $\alpha_2$ & $\beta_2$ \\
     \hline
     E1 & $100$ & $20$ & $4$ & $500$ & $1000$ & $20$ & $0.1$ & $500$\\
     E2 & $100$ & $20$ & $0.5$ & $500$ & $1000$ & $20$ & $0.5$ & $500$\\
     \hline
    \end{tabular}
    \caption{Parameter settings for the case of $G = 2$. Here, $n_g$ represents the number of hypotheses for the $g$th group; $n_{ga}$ denotes the number of non-null hypotheses in the $g$th group with $g=1,2$. $\alpha_g$ and $\beta_g$ are the parameters of the beta distribution for the p-values under the alternatives for the $g$th group.}
    \label{tab:fair-para}
\end{table}

\subsubsection*{Numerical Results for the Four-Group Setting}
We also consider the case of $G = 4$. To evaluate the performance of each method, we employ the following metrics: POW represents the overall power combining the rejections from all four groups; POW$_g$ denotes the power for the $g$th group with $1\leq g\leq 4$. Similarly, we can define FDR and FDR$_g$. The empirical power and FDR are computed based on 1,000 independent Monte Carlo simulations.

In all settings, we assume that the p-values follow the uniform distribution on $[0, 1]$ under the null. For the $g$th group, the p-value is assumed to follow Beta($\alpha_g, \beta_g$) under the alternatives. The parameter values for different settings are detailed in Table~\ref{tab:fair-para-appe}.

Table~\ref{tab:fair-appe1} displays the results for Setting F1. It can be seen that \texttt{BC\_Com} fails to control the FDR for the third and fourth groups, with the empirical FDR reaching 0.073 compared to the 5\% target level. \texttt{BC\_Sep} has an empirical FDR of 0.064, higher than the nominal level. The results for Setting F2 are presented in Table~\ref{tab:fair-appe2}. We note that \texttt{BC\_Com} suffers from a severe FDR inflation with the empirical FDR being 0.312 at the 5\% target level.  In Setting F3, we raise the target FDR level to $20\%$. As seen from Table~\ref{tab:fair-appe3}, \texttt{BC\_Sep} significantly inflates the overall FDR, with the empirical FDR being 0.346. 
Throughout all settings, the e-BH-based approach controls both the group-wise FDR and the overall FDR. Furthermore, \texttt{eBH\_Ada} outperforms both \texttt{eBH\_1} and \texttt{eBH\_2} in terms of power.

\begin{sidewaystable}
\centering
\begin{tabular}{|c|c|cccc|cccc|cccc|cccc|}
\hline
Setting & Target FDR level & $n_1$ & $n_{1a}$ & $\alpha_1$ & $\beta_1$ & $n_2$ & $n_{2a}$ & $\alpha_2$ & $\beta_2$ & $n_3$ & $n_{3a}$ & $\alpha_3$ & $\beta_3$ & $n_4$ & $n_{4a}$ & $\alpha_4$ & $\beta_4$\\
\hline
Setting F1 & 0.05 & 100 & 20 & $0.1$ & $500$ & 100 & 20 & $0.1$ & $500$ & 
1000 & 20 & $0.1$ & $500$ &  1000 & 20 & $0.1$ & $500$ \\
Setting F2 & 0.05 & 100 & 1 & $0.01$ & 5000 & 100 & 20 & $0.1$ & 500 & 
100 & 20 & $0.1$ & 500 &  100 & 20 & $0.1$ & $500$ \\
Setting F3 & 0.2 & 50 & 2 & $0.1$ & 500 & 100 & 2 & $0.1$ & 500 & 
50 & 4 & $0.2$ & & 500  100 & 4 & $0.3$ & 500 \\
\hline
\end{tabular}
\caption{Parameter settings for the case of $G = 4$. Here, $n_g$ represents the number of hypotheses for the $g$th group; $n_{ga}$ denotes the number of non-null hypotheses in the $g$th group. $\alpha_g$ and $\beta_g$ are parameters of the beta distribution for the p-values under the alternatives for the $g$th group.}
\label{tab:fair-para-appe}

\bigskip\bigskip\bigskip\bigskip\bigskip\bigskip

\begin{tabular}{|c|ccccc|ccccc|}
\hline
 Method & POW & POW$_1$ & POW$_2$ & POW$_3$ & POW$_4$ & FDR & FDR$_1$ & FDR$_2$ & FDR$_3$ & FDR$_4$ \\
\hline
\texttt{BC\_Com} & 0.954 & 0.955 & 0.956 & 0.952 & 0.953 & 0.045 & 0.007 & 0.006 & 0.078 & 0.077 \\
 \texttt{BC\_Sep} & 0.649 & 0.895 & 0.865 & 0.411 & 0.428 & 0.064 & 0.049 & 0.042 & 0.035 & 0.04 \\
\hline
 \texttt{eBH\_1} & 0.029 & 0 & 0 & 0.057 & 0.057 & 0.008 & 0 & 0 & 0.007 & 0.008 \\
 \texttt{eBH\_2} & 0.14 & 0.142 & 0.142 & 0.139 & 0.138 & 0.01 & 0.007 & 0.007 & 0.011 & 0.013 \\
 \texttt{eBH\_a} & 0.222 & 0.256 & 0.258 & 0.179 & 0.194 & 0.017 & 0.013 & 0.012 & 0.015 & 0.018 \\
\hline
\end{tabular}
\caption{FDR and power for Setting F1, where the nominal FDR level is 5\%.}
\label{tab:fair-appe1}
\end{sidewaystable}

\begin{sidewaystable}
\centering
\begin{tabular}{|c|ccccc|ccccc|}
\hline
 Method & POW & POW$_1$ & POW$_2$ & POW$_3$ & POW$_4$ & FDR & FDR$_1$ & FDR$_2$ & FDR$_3$ & FDR$_4$ \\
\hline
\texttt{BC\_Com} & 0.988 & 0.999 & 0.994 & 0.987 & 0.983 & 0.046 & 0.318 & 0.034 & 0.033 & 0.032 \\
\texttt{BC\_Sep} & 0.793 & 0 & 0.863 & 0.784 & 0.771 & 0.055 & 0 & 0.045 & 0.043 & 0.042 \\
\hline
\texttt{eBH\_1} & 0 & 0 & 0 & 0 & 0 & 0 & 0 & 0 & 0 & 0 \\
\texttt{eBH\_2} & 0 & 0 & 0 & 0 & 0 & 0 & 0 & 0 & 0 & 0 \\
\texttt{eBH\_a} & 0.53 & 0 & 0.544 & 0.537 & 0.536 & 0.031 & 0 & 0.028 & 0.03 & 0.03 \\
\hline
\end{tabular}
\caption{FDR and power for Setting F2, where the nominal FDR level is 5\%.}
\label{tab:fair-appe2}

\bigskip\bigskip\bigskip\bigskip\bigskip\bigskip

\begin{tabular}{|c|ccccc|ccccc|}
\hline
 Method & POW & POW$_1$ & POW$_2$ & POW$_3$ & POW$_4$ & FDR & FDR$_1$ & FDR$_2$ & FDR$_3$ & FDR$_4$ \\
\hline
\texttt{BC\_Com} & 0.983 & 0.992 & 0.991 & 0.984 & 0.975 & 0.181 & 0.154 & 0.249 & 0.089 & 0.161 \\
\texttt{BC\_Sep} & 0.393 & 0.149 & 0.141 & 0.517 & 0.517 & 0.343 & 0.1 & 0.094 & 0.164 & 0.176 \\
\hline
\texttt{eBH\_1} & 0.002 & 0 & 0.003 & 0 & 0.003 & 0.002 & 0 & 0.002 & 0 & 0.001 \\
\texttt{eBH\_2} & 0.004 & 0.005 & 0 & 0.005 & 0.005 & 0.003 & 0.003 & 0 & 0.001 & 0.002 \\
\texttt{eBH\_a} & 0.035 & 0.013 & 0.006 & 0.049 & 0.047 & 0.019 & 0.008 & 0.004 & 0.016 & 0.015 \\
\hline
\end{tabular}
\caption{FDR and power for Setting F3, where the nominal FDR level is 20\%.}
\label{tab:fair-appe3}
\end{sidewaystable}

\subsection{Numerical Studies for the Real Dataset}\label{appe:fair-real}
We illustrate the proposed method by conducting differential abundance analysis using the microbiome dataset \texttt{cdi\_schubert}, sourced from the MicrobiomeHD repository \citet{duvallet2017microbiomehd}, originally collected in a case-control study comparing individuals with Clostridium difficile infection (CDI) to those without (nonCDI) \citet{schubert2014microbiome}. This dataset comprises 336 microbiome samples. Each sample is classified as either CDI (infection case) or non-CDI (healthy control). The raw feature table contains a total of 19,314 operational taxonomic units (OTUs), representing bacterial taxa annotated at the phylum level.

Before analysis, several filtering and preprocessing steps were applied to refine the OTU table. Initially, OTUs lacking phylum annotations were excluded. Subsequently, entire phylum groups containing fewer than 200 taxa were removed, restricting analysis to well-represented bacterial groups. Following this step, features from three major bacterial phyla remained: Bacteroidetes, Firmicutes, and Proteobacteria. Additionally, we implemented prevalence-based quality control by filtering out taxa detected in fewer than 10 samples. After applying these criteria, the final dataset retained 2293 microbiome features across all 336 samples, ensuring robust and informative taxa for downstream analyses.

We then performed differential abundance testing to identify taxa differing between the diarrheal case and control groups. Specifically, we utilized the LinDA method \citep{zhou2022linda}, which fits a log-linear model to compositional microbiome data, adjusting for sequencing depth and compositional bias. This method generated a p-value for each taxon, assessing differences in abundance between cases and controls. The resulting collection of p-values was subsequently processed using multiple-testing correction procedures. In particular, we applied the \texttt{BC\_Com} method and three eBH-based methods (\texttt{eBH\_1}, \texttt{eBH\_2}, and \texttt{eBH\_Ada}), as proposed before, with a target FDR of $\alpha = 0.2$. The number of rejected taxa for each phylum is summarized in Table~\ref{tab:method_phylum}.

\begin{table}
\centering
\begin{tabular}{|l|ccc|}
\hline
\textbf{method/phylum} &  Bacteroidetes & Firmicutes & Proteobacteria  \\
\hline
\texttt{BC\_Com} & 354  & 515 & 106  \\
\texttt{eBH\_1}  & 0 & 0 & 0 \\
\texttt{eBH\_2}  & 0 & 0 & 0 \\
\texttt{eBH\_Ada}  &  259  & 557 & 175  \\
\hline
\end{tabular}
\caption{Numbers for rejections for each method and phylum.}
\label{tab:method_phylum}
\end{table}

In this study, controlling the overall FDR ensures the reliability of global inference, whereas controlling the FDR within each phylum is essential for accurately interpreting results within biologically meaningful groups. The results in Table~\ref{tab:method_phylum} show that for the phyla Bacteroidetes and Firmicutes, the \texttt{eBH\_Ada} method yields fewer rejections compared to the \texttt{BC\_Com} method, suggesting that \texttt{BC\_Com} might inadequately control FDR within these specific groups. Conversely, for the phylum Proteobacteria, the \texttt{eBH\_Ada} methods identify a greater number of rejections, mirroring the pattern observed in our simulation scenario E1, where the \texttt{BC\_Com} method exhibits reduced power in certain groups. Additionally, the two data-independent weighting methods, \texttt{eBH\_1} and \texttt{eBH\_2}, have no discovery, underscoring the practical necessity of employing data-dependent weights when combining e-values.

\newpage
\section{Hybrid Knockoff Procedure} 
In this section, we demonstrate how e-values can be used to combine results obtained from different test statistics. The knockoff method \citep{barber2015controlling, candes2018panning} provides a variable selection framework designed to control the FDR at a specified level. Typically, knockoff methods utilize the Lasso coefficient-difference statistic; however, when the relationship between response and regressors is non-linear, statistics based on random forests become preferable \citep{candes2018panning}. Given that the true dependence structure between the response and regressors is usually unknown in practice, we propose a hybrid knockoff approach that combines results from multiple test statistics, enhancing robustness across different modeling scenarios. Specifically, motivated by the recent work of \citet{ren2024derandomised}, which establishes the equivalence between the knockoff procedure and the e-BH procedure under certain e-values, our approach first transforms results from multiple knockoff statistics into corresponding e-values. These e-values are then aggregated via arithmetic mean, and the e-BH procedure is subsequently applied to determine the final rejection set. A detailed exposition of the knockoff framework and the proposed hybrid method is provided in the subsequent sections.

\subsection{Connection between Knockoff and e-BH Procedures}
In \citet{ren2024derandomised}, the authors demonstrated that the knockoff method is equivalent to the e-BH method under a specific form of e-values. In this subsection, we briefly review this result for later use.

In variable selection, the goal is to identify predictors that are significantly associated with the response variable. A predictor is considered a null variable if it is conditionally independent of the response given all other predictors. Formally, let $Y$ denote the response variable. For a predictor \(X_j\), with the remaining predictors denoted by \(X_{-j} = \{X_i : 1 \leq i \leq p, \, i \neq j\}\), \(X_j\) is a null variable if  
\[Y \indep X_j \mid X_{-j}.\]

Now, consider the linear model  
\[\mathbf{Y} = \mathbf{X} \boldsymbol{\beta} + \boldsymbol{\epsilon},\]
where \(\mathbf{Y} \in \mathbb{R}^n\) is the response vector, \(\mathbf{X} = [\bfX_1, \ldots, \bfX_p] \in \mathbb{R}^{n \times p}\) is the covariate matrix with \(\bfX_j\) as its \(j\)th column, and \(\boldsymbol{\beta} = [\beta_1, \dots, \beta_p]^\top \in \mathbb{R}^p\) is the vector of regression coefficients. Suppose that the error term \(\boldsymbol{\epsilon} \in \mathbb{R}^n\) follows the normal distribution $\mathcal{N}(0, \sigma^2 \mathbf{I}_n),$ and is independent of \(\mathbf{X}\). Under this framework, testing the hypothesis \(Y \indep X_j \mid X_{-j}\) is equivalent to testing whether \(\beta_j = 0\). In practice, if a variable selection procedure returns a set of indices \(\hat{S} \subset \{1, 2, \ldots, p\}\), the FDR is defined as  
\[\text{FDR} = \mathbb{E}\left[\frac{\#\{i:\beta_i = 0,\, i \in \hat{S}\}}{1\vee|\hat{S}|}\right].\]

The knockoff method \citep{barber2015controlling, candes2018panning} provides a variable selection approach that controls the FDR at a desired level when \(n > p\). The key idea is to construct synthetic predictors, \(\tilde{\mathbf{X}}\), called knockoffs, which preserve the correlation structure of the original features \(\mathbf{X}\) while remaining conditionally independent of the response \(\mathbf{Y}\). Specifically, the knockoff procedure generates a knockoff copy \(\tilde{\mathbf{X}}\) that satisfies  
\[\mathbf{Y} \indep \tilde{\mathbf{X}} \mid \mathbf{X}\]
and  
\[(\bfX_j, \tilde{\bfX}_j, \bfX_{-j}, \tilde{\bfX}_{-j}) \overset{d}{=} (\tilde{\bfX}_j, \bfX_j, \tilde{\bfX}_{-j}, \bfX_{-j}),\]
for each \(j\), where \(\overset{d}{=}\) denotes equality in distribution. For details on constructing \(\tilde{\mathbf{X}}\), see \citet{barber2015controlling, candes2018panning, barber2020robust}.

Once the knockoffs \(\tilde{\mathbf{X}}\) are constructed, feature importance statistics are computed using the augmented data \(([\tilde{\mathbf{X}}, \mathbf{X}], Y)\) by  
\[\bfW = \mathcal{W}([\tilde{\mathbf{X}}, \mathbf{X}], \mathbf{Y}),\]
where \(\mathcal{W}(\cdot)\) is an algorithm that quantifies the importance of each feature. A key property of \(\bfW = [W_1, \dots, W_p]^\top\) is that swapping \(\bfX_j\) and \(\tilde{\bfX}_j\) reverses the sign of \(W_j\), and larger values of \(W_j\) provide stronger evidence against the null hypothesis for the \(j\)th predictor. For a predetermined FDR level \(\alpha\), the knockoff threshold \(T\) is defined as  
\[T = \min\left\{t \in \mathbb{W} \colon \frac{1 + \sum_{j=1}^p \ind\{W_j \leq -t\}}{1 \vee \sum_{j=1}^p \ind\{W_j \geq t\}} \leq \alpha \right\},\]
where \(\mathbb{W} = \{|W_j| \colon i = 1, 2, \ldots, p\} \setminus \{0\}\). The final selected model is given by  
\[\hat{S} = \{j : W_j \geq T\}.\]
The FDR control properties of the knockoff method are detailed in \citet{barber2015controlling, barber2020robust}.

In \citet{ren2024derandomised}, the authors demonstrated that the e-BH procedure, using the e-value defined for the $i$th hypothesis as
\begin{equation}\label{eq:e-value-knockoff}
e_i \coloneqq \frac{p \ind\{W_i \geq T\}}{1 + \sum_{j=1}^p \ind\{W_j \leq -T\}},
\end{equation}
is equivalent to the knockoff method. Specifically, the two procedures yield identical rejection sets.

\subsection{Hybrid Knockoff Algorithm}
In knockoff methods, various test statistics can be utilized, and the optimal choice generally depends on the underlying data-generating mechanism. Since the true relationship between $Y$ and $X$ is unknown in practice, selecting a suitable test statistic in advance is challenging. In this section, we propose a hybrid knockoff approach that combines results from different test statistics, ensuring robust performance under various scenarios.

One of the most commonly used test statistics is the Lasso coefficient-difference statistic proposed by \citet{candes2018panning}. Specifically, one fits a cross-validated Lasso regression to the augmented design matrix $[\bfX, \tilde{\bfX}]$ to predict $\bfY$. Denoting by $\beta_j$ and $\tilde{\beta}_j$ the fitted coefficients corresponding to the original feature $\bfX_j$ and its knockoff $\tilde{\bfX}_j$, respectively, the test statistic is defined as 
\[W_j = |\beta_j| - |\tilde{\beta}_j|,\quad j = 1,\dots,p.\]
Intuitively, this statistic performs well when the true relationship between $\bfY$ and $\bfX$ is approximately linear \citep{candes2018panning}. However, if this linear assumption does not hold, alternative statistics based on nonlinear models, such as random forests, can be considered. For example, one can define $W_j$ as the difference in feature importance scores between the original feature $\bfX_j$ and its knockoff counterpart $\tilde{\bfX}_j$ computed by a random forest model \citep{candes2018panning}.

Since we do not know the true dependence structure between $\bfY$ and $\bfX$, it is beneficial to develop a unified procedure that performs consistently well regardless of the underlying model form. Leveraging the concept of e-values, we propose combining results from multiple knockoff test statistics into a single robust procedure. Specifically, we first compute knockoff statistics separately, using both Lasso-based and random-forest-based approaches, and then transform these statistics into e-values. These individual e-values are subsequently combined into a single e-value vector, after which the e-BH procedure is applied to determine the final rejection set.

Formally, let $\bfW_1 = [W_{1,1}, \dots, W_{1,p}]^\top$ and $\bfW_2 = [W_{2,1}, \dots, W_{2,p}]^\top$ denote two distinct sets of knockoff test statistics. First, we implement the knockoff methods using these two sets of statistics at the target FDR level $\alpha_{\mathrm{ko}}$, obtaining thresholds $T_1$ and $T_2$. Following the discussion in \citet{ren2024derandomised}, a standard choice is $\alpha_{\mathrm{ko}} = \alpha_{\ebh} / 2$, where $\alpha$ is the target FDR level used in the e-BH procedure. Subsequently, we apply equation~\eqref{eq:e-value-knockoff} to compute individual e-value vectors $\bfe_1$ and $\bfe_2$ based on $(\bfW_1, T_1)$ and $(\bfW_2, T_2)$, respectively. By construction, these e-value vectors satisfy 
\[\sum_{i\in\cH_0} \bbE[\bfe_{1,i}] \leq p, \quad \sum_{i\in\cH_0} \bbE[\bfe_{2,i}] \leq p.\]
We propose combining the two vectors into a unified e-value vector $\bfe$, defined component-wise as
\begin{equation}\label{eq:com-e-knockoff}
\bfe_i = w_1 \bfe_{1,i} + w_2 \bfe_{2,i}, \quad \text{with } w_1 + w_2 \leq 1.
\end{equation}
This ensures that the resulting e-value vector $\bfe$ also satisfies the condition given by \eqref{eq:evalue}. A natural default choice is $w_1 = w_2 = 0.5$. By Proposition~\ref{prop:ebh-control}, the proposed approach controls the FDR at the desired level. Algorithm~\ref{alg:hybrid-knockoff} describes the detailed implementation of this procedure.

\begin{algorithm}  
\caption{Hybrid Knockoff Procedure}\label{alg:hybrid-knockoff}
\begin{algorithmic}[1]
\Require Response vector $\bfY$, covariate matrix $\bfX$, and significance level $\alpha_{\ebh}$.
\State Run the knockoff procedure with Lasso-based test statistics at target FDR level $\alpha_{\ebh} / 2$ to obtain test statistic vector $\bfW_1$ and threshold $T_1$. Compute the corresponding e-values:
\[\bfe_{1,i} = \frac{n\,\ind\{W_{1,i} \geq T_1\}}{1 + \sum_{j=1}^n\ind\{W_{1,j}\leq -T_1\}}, \quad i=1,\dots,n.\]
\State Run the knockoff procedure with random-forest-based test statistics at target FDR level $\alpha_{\ebh} / 2$ to obtain test statistic vector $\bfW_2$ and threshold $T_2$. Compute the corresponding e-values:
\[\bfe_{2,i} = \frac{n\,\ind\{W_{2,i} \geq T_2\}}{1 + \sum_{j=1}^n\ind\{W_{2,j}\leq -T_2\}}, \quad i=1,\dots,n.\]
\State Compute the combined weighted average e-values according to equation~\eqref{eq:com-e-knockoff}.
\State Apply the e-BH procedure to the combined e-values at the significance level $\alpha_{\ebh}$.
\Ensure Indices of the rejected hypotheses.
\end{algorithmic}
\end{algorithm}

\subsection{Numerical Studies for the Hybrid Knockoff Method}
We adopt the simulation settings from \citet{ren2024derandomised}. In the first scenario, we generate data from a Gaussian linear model. Specifically, we fix the target FDR level $\alpha = 0.2$, the sample size at $n = 1,000$ and the feature dimension at $p = 800$, while varying the signal sparsity $|\cH_1| \in \{40, 80\}$, where $\cH_1$ is the collection of alternative hypotheses.

In this first scenario, the covariate matrix $\bfX$ is drawn from a multivariate normal distribution $\cN(\mathbf{0}, \Sigma)$, where the covariance matrix $\Sigma$ is defined by $\Sigma_{ij} = 0.5^{|i - j|}$. The response vector $\bfY$ is generated according to the linear model $\bfY \sim \cN\bigl(\bfX\bsbeta, 1\bigr)$, where the nonzero regression coefficients $\beta_i$ are sampled independently from $\beta_i \sim \cN(\mu / \sqrt{n}, 1)$, with the mean parameter $\mu$ controlling the signal strength and varying within the interval $[2.5, 4]$. Additionally, half of the nonzero coefficients $\beta_i$ are set to be positive, and the other half are set to be negative.

In the second scenario, we retain the sample size, feature dimension, and signal sparsity. We generate the response vector $\mathbf{Y}$ according to a nonlinear model. For any matrix $\mathbf{X}$, let $\mathbf{X}_{l_1:l_2}$ denote the submatrix comprising columns from index $l_1$ to index $l_2$. Similarly, for any vector $\boldsymbol{\beta}$, let $\boldsymbol{\beta}_{l_1:l_2}$ denote the subvector containing elements from index $l_1$ to index $l_2$. For any function $f$, $f(\mathbf{X})$ means that $f$ is applied to each component of $\mathbf{X}$. We generate the data using the following nonlinear model:
\[\mathbf{Y} \sim \mathcal{N}\left(\sin(\mathbf{X}_{1:p/2})\,\boldsymbol{\beta}_{1:p/2} + (\mathbf{X}_{(p/2+1):p})^2\,\boldsymbol{\beta}_{(p/2+1):p},\,1\right).\]
Thus, the relationship between the response $\mathbf{Y}$ and the first $p/2$ columns of $\mathbf{X}$ involves a sine transformation, while the last $p/2$ columns of $\mathbf{X}$ are squared. This creates a nonlinear dependence between $\mathbf{Y}$ and $\mathbf{X}$. Similar to the linear scenario, the covariance matrix $\Sigma$ used to generate covariate matrix $\bfX$ is defined by $\Sigma_{ij} = 0.1^{|i - j|}$, the nonzero regression coefficients $\beta_i$ are independently sampled from $\beta_i \sim \cN(\mu / \sqrt{n}, 1)$, but now with the mean parameter $\mu$ varying within the interval $[5, 20]$. Again, half of the nonzero coefficients $\beta_i$ are set to be positive and half negative.

\begin{figure}
    \centering
    \includegraphics[width=0.8\linewidth]{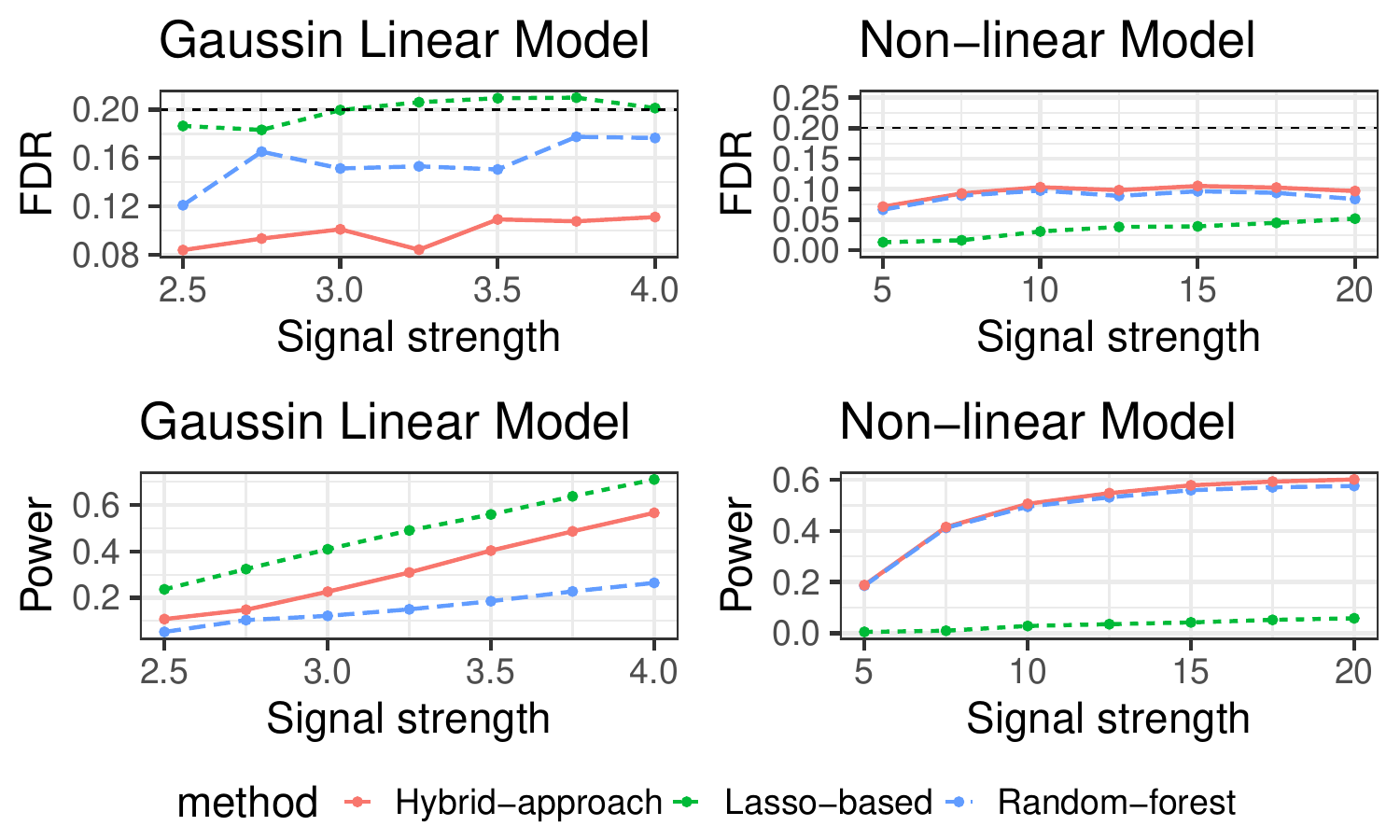}
    \caption{Simulation results for hybrid knockoff methods under two scenarios: Gaussian linear model (left panel) and nonlinear model (right panel).}
    \label{fig:knock-off-40}
\end{figure}

\begin{figure}
    \centering
    \includegraphics[width=0.8\linewidth]{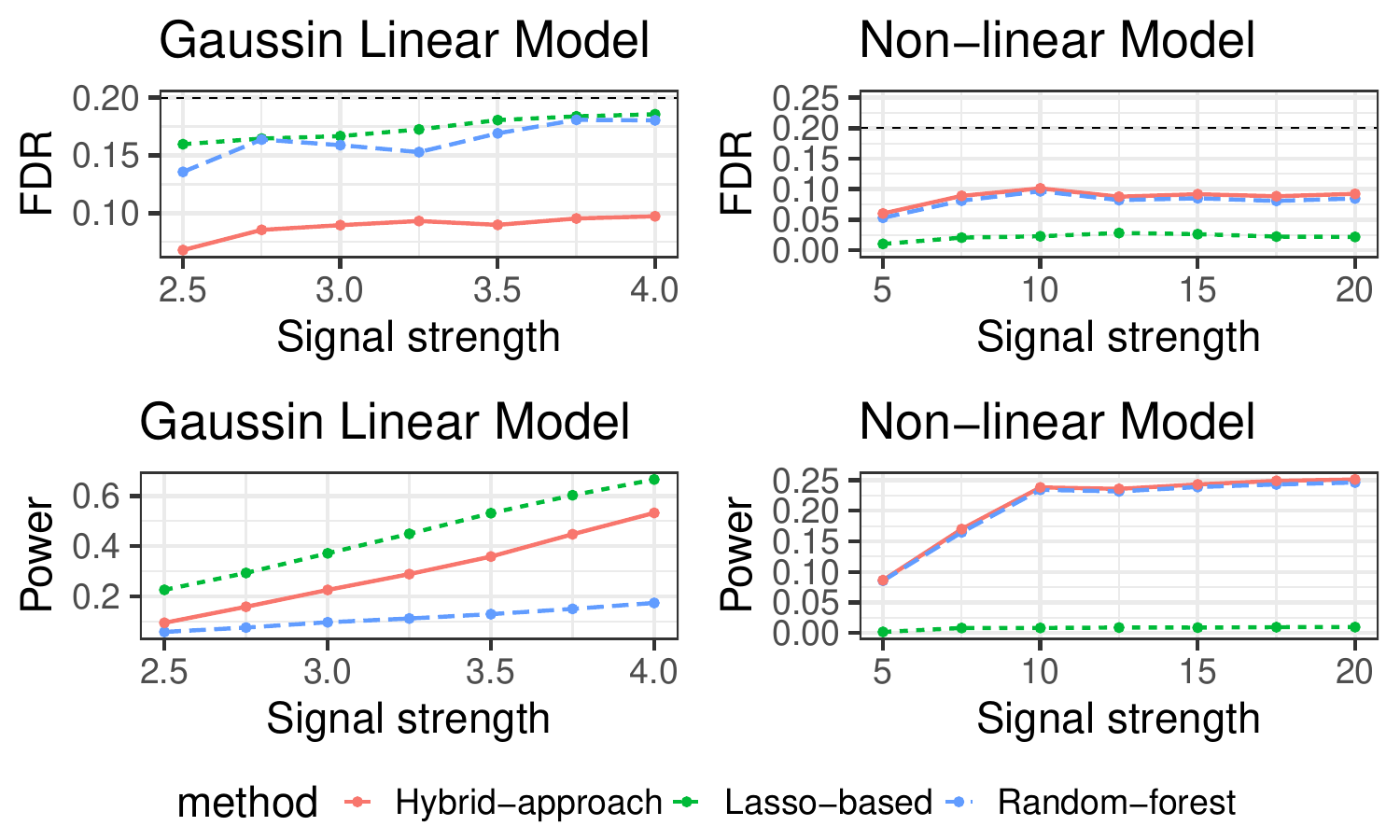}
    \caption{Simulation results for hybrid knockoff methods under two scenarios: Gaussian linear model (left panel) and nonlinear model (right panel).}
    \label{fig:knock-off-80}
\end{figure}

The results under both scenarios when $|\cH_1| = 40$ are summarized in Figure~\ref{fig:knock-off-40}. The case with $|\cH_1|=80$ exhibits a similar pattern; we have included these results in the Figure~\ref{fig:knock-off-80}.  Across all simulation settings, almost every method controls the false discovery rate; only the Lasso-based procedure exhibits slight FDR inflation under the linear-model scenario. Regarding the power, under the linear model scenario, the Lasso-based method outperforms the random-forest-based method, whereas in the nonlinear scenario, the random-forest-based method exhibits superior performance. Notably, our proposed hybrid approach consistently outperforms the weaker method of the two and, in several instances in the non-linear model case, surpasses both. Thus, the proposed method effectively integrates the advantages of both methods and demonstrates robustness across diverse model settings. 

\newpage
\section{Additional Details for the Hybrid Multiple Testing Procedure}
\subsection{The Way to Generate P-values}
We generate p-values as follows. Under the null hypothesis, we simulate the test statistics $X_i$ from the standard normal distribution $\cN(0, 1)$. Under the alternative hypothesis, $X_i$ follows $\cN(\mu\log(n), \sigma^2)$, where the parameter $\mu$ governs the signal strength. The p-values are then computed as $p_i = 1 - \Phi(X_i)$, where $\Phi$ denotes the cumulative distribution function of the standard normal distribution.

In Setting S1, we fix the sample size at $n = 1,000$ and set $n_a = 50$, corresponding to $5\%$ signals. The signal strength $\mu$ varies between $0.3$ and $0.5$, and $\sigma = 1$. In Setting S2, we fix the sample size at $n = 3,000$ and set $n_a = 750$, corresponding to $25\%$ signals. The signal strength $\mu$ varies between $0.275$ and $0.295$, and $\sigma = 0.4$. 

\subsection{The Computational Cost for the Hybrid Approach}
The computational cost for Setting S1 with $\mu = 0.4$ for each method are shown in Table~\ref{tab:hyper-time}.
\begin{table}
    \centering
    \begin{tabular}{|c|cccccc|}
    \hline
    method & BH & BC & ST & \texttt{eBH\_Ave} & \texttt{eBH\_Ada} & \texttt{fast\_eBH\_Ada} \\
    \hline
    time in seconds  & $1\times 10^{-4}$ & $2\times 10^{-4}$& $1\times 10^{-4}$ & $2\times 10^{-4}$&2.9 & 0.17\\
    \hline
    \end{tabular}
    \caption{The average running time for each method in the 500 simulations. \texttt{eBH\_Ave} denotes the hybrid approach with data-independent weights. \texttt{eBH\_Ada} denotes the hybrid approach with data-dependent weights, and \texttt{fast\_eBH\_Ada} is the fast version of the \texttt{eBH\_Ada} procedure.}
    \label{tab:hyper-time}
\end{table}

\section{Additional Details for Structure-Adaptive Multiple Testing}
\subsection{Structure-Adaptive Multiple Testing via Cross-Fitting}
Algorithm~\ref{alg:e-gbc-stru} summarizes the structure-adaptive multiple testing procedure via cross-fitting.
\begin{algorithm}
\caption{Cross-fitting based structure adaptive multiple testing}\label{alg:e-gbc-stru}
\begin{algorithmic}[1]
\Require p-values $p_1, \dots, p_n$, covariates $x_1, \dots, x_n$, group indices $\cG_1, \dots, \cG_G$, significance levels $\alpha_{\fbc}$ and $\alpha_{\ebh}$
\For{$g = 1, \dots, G$}
\State Compute the rejection function estimate $\hat \varphi_{i}(\cdot)$ for $i\in\cG_g$ using $\bfp_{-g}$ and $\bfx_{-g}$.
\State Calculate the threshold $T_g$ using \eqref{eq:crossthres-gbc}.
\For{$i\in\cG_g$}
\State Calculate the e-value $e_{i}$ using \eqref{eq:cross-evalue-gbc}.
\EndFor 
\EndFor
\State Assemble the e-values from all groups.
\State Run the e-BH procedure utilizing the assembled e-values at the level $\alpha_{\ebh}$.
\Ensure The indices of rejected hypotheses.
\end{algorithmic}
\end{algorithm}

\subsubsection*{Example: Data-Dependent Weights}
% {\color{red} Maybe add one or two sentences here. one or two sentences here.}
Below is a concrete example illustrating the form of the data-dependent weights. In the case of $G = 2$, we have
\[w_i = \frac{\frac{n}{n_1}\left(1 + \sum_{j\neq i, j\in\cG_1}\ind\{\hat \varphi_{j}(1 - p_{j})\leq T_1\}\right)}{\left(1 + \sum_{j\neq i, j\in\cG_1}\ind\{\hat \varphi_{j}(1 - p_{j})\leq T_1\}\right) + \sup_{p\in[0, 1]}\sum_{j\in\cG_2}\ind\{\hat \varphi_{j}^{1, i, p}(1 - p_{j}) \leq T_{2, j}^{1, i, p}\}}\]
for $i\in\cG_1$ and
\[w_i = \frac{\frac{n}{n_2}\left(1 + \sum_{j\neq i, j\in\cG_2}\ind\{\hat \varphi_{j}(1 - p_{j})\leq T_2\}\right)}{\sup_{p\in[0, 1]} \sum_{j\in\cG_1}\ind\{\hat \varphi_{j}^{2, i, p}(1 - p_{j}) \leq T_{1, j}^{2, i, p}\} + \left(1 + \sum_{j\neq i, j\in\cG_2}\ind\{\hat \varphi_{j}(1 - p_{j})\leq T_2\}\right)}\]
for $i\in\cG_2$. 

\subsubsection*{Introduction of Local FDR}    
In the FBC procedure, we employ a rejection rule based on the local FDR \citep{sun2007oracle} within the two-group mixture model framework. Specifically, assume that the p-value $p_i$ is independently generated from the two-group mixture model: $\pi_i f_0 + (1 - \pi_i) f_{1,i}$, where $\pi_i \in (0, 1)$ is the mixing proportion, and $f_{0}$ and $f_{1,i}$ represent the p-value distributions under the null and alternative hypotheses, respectively. The local FDR is defined as
\[\lfdr_i(p) = \frac{\pi_i f_0(p)}{\pi_i f_0(p) + (1 - \pi_i) f_{1,i}(p)},\]
which represents the posterior probability that the $i$th hypothesis is null given the observed p-value $p$. The monotone likelihood ratio assumption \citep{sun2007oracle} posits that $f_{1,i}(p) / f_0(p)$ is decreasing in $p$. Under this assumption, $\varphi_i(p) = \lfdr_i(p)$ is monotonically increasing in $p$, which satisfies the conditions for the FBC procedure to control FDR at the target level \citep{li2025note}. Furthermore, the literature demonstrates that the rejection rule $\ind\{\varphi_i(p_i) = \lfdr_i(p_i) \leq t\}$ is optimal in maximizing the expected number of true positives among decision rules that control the marginal FDR at level $\alpha$ (see, e.g., \cite{sun2007oracle, lei2018adapt, cao2022optimal}).

\subsubsection*{Parameter Estimation}
In practice, we using Local FDR as the hypothesis specific rejection function. Set $f_0(p)=\ind\{p\in [0,1]\}$ and $f_{1,i}(p)=(1 - \pi_i)(1-\kappa_i)p^{-\kappa_i}$ with $\kappa_i\in (0,1).$ We consider the working models that link $(\pi_i,\kappa_i)$ with the external covariates:
\begin{align*}
\pi_{i} &= \pi_{\beta_{\pi}}(x_i) = \frac{1}{1 + \exp(-\beta_{\pi, 0} - \beta_{\pi, 1}^\top x_i)}, \\
\kappa_i &= \kappa_{\beta_{\kappa}}(x_i) = \frac{1}{1 + \exp(-\beta_{\kappa, 0} - \beta_{\kappa, 1}^\top x_i)},  
\end{align*}
where the parameters $\beta_{\pi}=(\beta_{\pi,0},\beta_{\pi,1}^\top)\in\mathbb{R}^{d+1}$ and $\beta_{\kappa}=(\beta_{\kappa,0},\beta_{\kappa,1}^\top)\in\mathbb{R}^{d+1}$ can be estimated by maximizing the pseudo-log-likelihood using the EM algorithm. Please refer to \citet{zhang2022covariate} for more optimization details. After obtaining the estimates  $\hat{\beta}_{\pi}$ and $\hat{\beta}_{\kappa}$ from the EM algorithm, we define
\[\hat \pi_{i} = 
\begin{cases}
\epsilon_1 & \text{if }1 / \bigl(1 + \exp(-\hat \beta_{\pi, 0} - \hat \beta_{\pi, 1}^\top x_i)\bigr) \leq \epsilon_1,\\
1 / \bigl(1 + \exp(-\hat \beta_{\pi, 0} - \hat \beta_{\pi, 1}^\top x_i)\bigr) & \text{if }\epsilon_1 < 1 / \bigl(1 + \exp(-\hat \beta_{\pi, 0} - \hat \beta_{\pi, 1}^\top x_i)\bigr) < 1-\epsilon_2,\\
1 - \epsilon_2 & \text{otherwise,}
\end{cases}\]
where winsorization is used to prevent $\hat{\pi}_i$ from being too close to zero or one to stabilize the algorithm. We define the rejection rule 
\[\hat \varphi_i(p) = \frac{\hat \pi_i}{\hat{\pi}_i + (1 - \hat{\pi}_i)(1 - \hat{\kappa}_i)p^{-\hat{\kappa}_i}}\leq t,\]
where $\hat{\kappa}_i = 1 / \bigl(1 + \exp(-\hat\beta_{\kappa, 0} - \hat \beta_{\kappa, 1}^\top x_i)\bigr)$. We then apply the FBC procedure with the estimated $\hat{\varphi}_i$ at the target FDR level $\alpha$. The corresponding e-values are computed via \eqref{eq:cross-evalue-gbc}, and the weights are obtained from \eqref{eq:weight-struc}. To reduce computational cost, we introduce the following, less expensive weighting scheme:
\[w_{i} = \frac{\frac{n}{n_g}\left(1 + \sum_{j\neq i, j\in\cG_g}\ind\{\hat \varphi_{j}(1 - p_{j})\leq T_g\}\right)}{\left(1 + \sum_{j\neq i, j\in\cG_g}\ind\{\hat \varphi_{j}(1 - p_{j})\leq T_g\}\right) + \sum_{g'\neq g}\sum_{j\in\cG_{g'}}\ind\{\hat \varphi_{j}(1 - p_{j}) \leq T_{g', j}\}}.\]

In practice, we fix $\epsilon_1 = 0.1$, $\epsilon_2 = 1 \times 10^{-5}$, $G = 2$ with $|\mathcal{G}_1|=|\mathcal{G}_2|$, and $\alpha_{\fbc} = \alpha_{\ebh} / (1 + \alpha_{\ebh})$.

\subsubsection*{Benchmark Methods}
\begin{itemize}
    \item \texttt{BH}: The BH procedure \citep{benjamini1995controlling}. We implement this method using the \texttt{p.adjust} function in \texttt{R}.
    
    \item \texttt{IHW\_storey}: The covariate-powered cross-weighted method with Storey's procedure to estimate the null-proportion \citep{ignatiadis2021covariate}. We implement this method using the \texttt{ihw\_bh} function in the \texttt{R} package \texttt{IHWStatsPaper}.
    
    \item \texttt{IHW\_betamix}: The covariate-powered cross-weighted method with the beta mixture model \citep{ignatiadis2021covariate}. We implement this method using the \texttt{ihw\_betamix\_censored} function in the \texttt{R} package \texttt{IHWStatsPaper}.
    
    \item \texttt{AdaPT}: The adaptive p-value thresholding procedure \citep{lei2018adapt}. We implement this method using the \texttt{adapt\_glm} function in the \texttt{R} package \texttt{adaptMT}.
    
    \item \texttt{SABHA}: The structure adaptive BH procedure \citep{li2019multiple}. The code was downloaded from the link provided by the original paper.
\end{itemize}

\subsubsection*{Comparison of Proposed Method with $\tau$-censored Weighted BH Procedure}
The $\tau$-censored weighted BH procedure proposed by \cite{zhao2024tau} is essentially a variant of the weighted BH procedure that uses a leave-one-out technique to construct weights. Our method is built upon the BC procedure, and the weights in our approach are for combining the e-values from different groups (obtained through sample-splitting). In other words, the weights serve different goals in the two procedures. The exact constructions of the weights in the two procedures are also very different. However, the two methods do use a similar trick to construct weights. In \cite{zhao2024tau}, the weight involves taking an infimum over $p_i\in [0,1]$ in the initial weights (see Section 3.1 therein). In our procedure, the construction of the weights involves taking the supremum over $p_i\in[0,1]$, which is crucial for the proof to go through.

\subsection{Additional Numerical Results for Structure-Adaptive Multiple Testing}\label{appe:ebh-gbh}
\begin{figure}
    \centering
    \includegraphics[width = 0.95\textwidth]{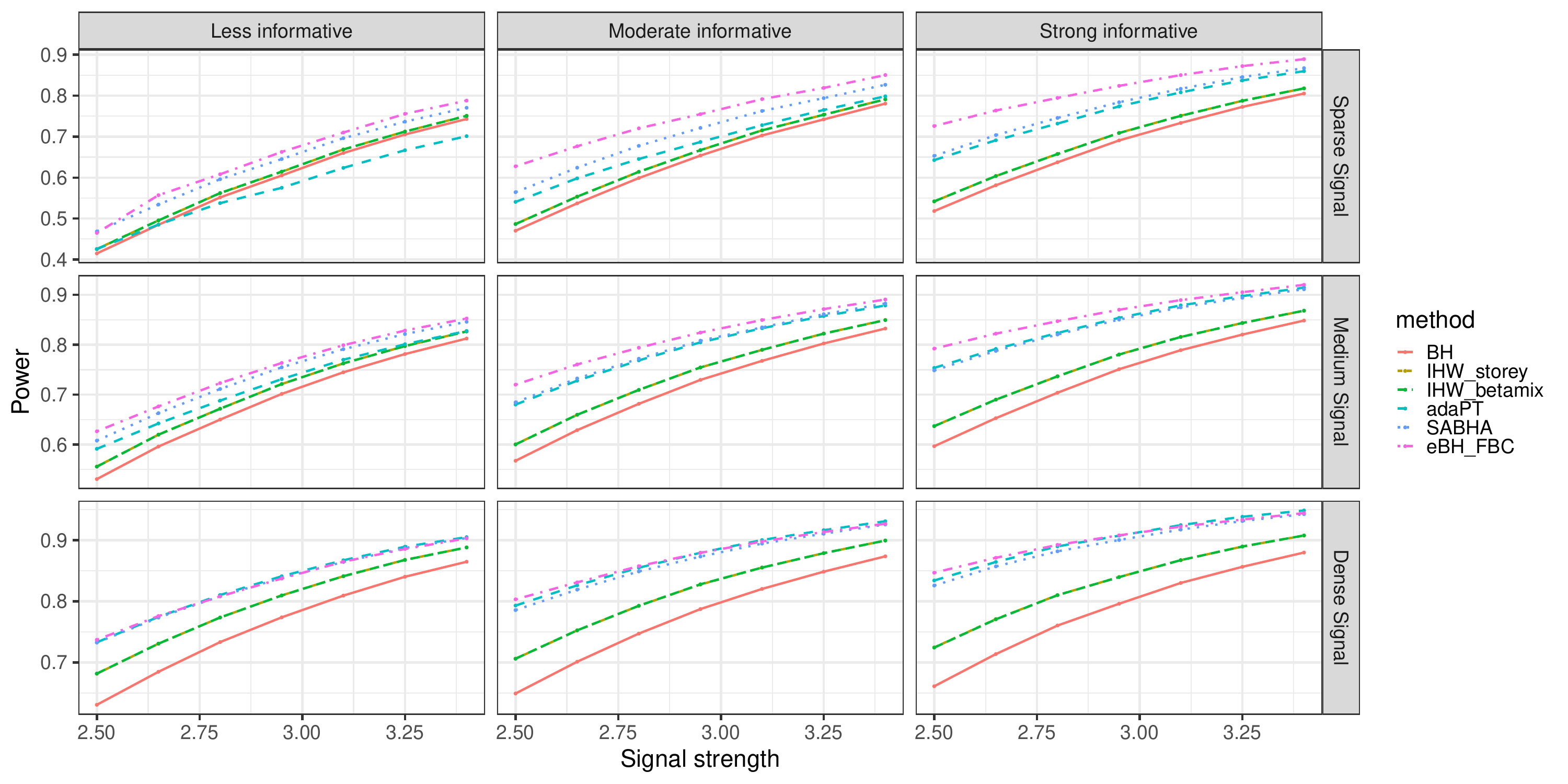}\\
    \includegraphics[width = 0.95\textwidth]{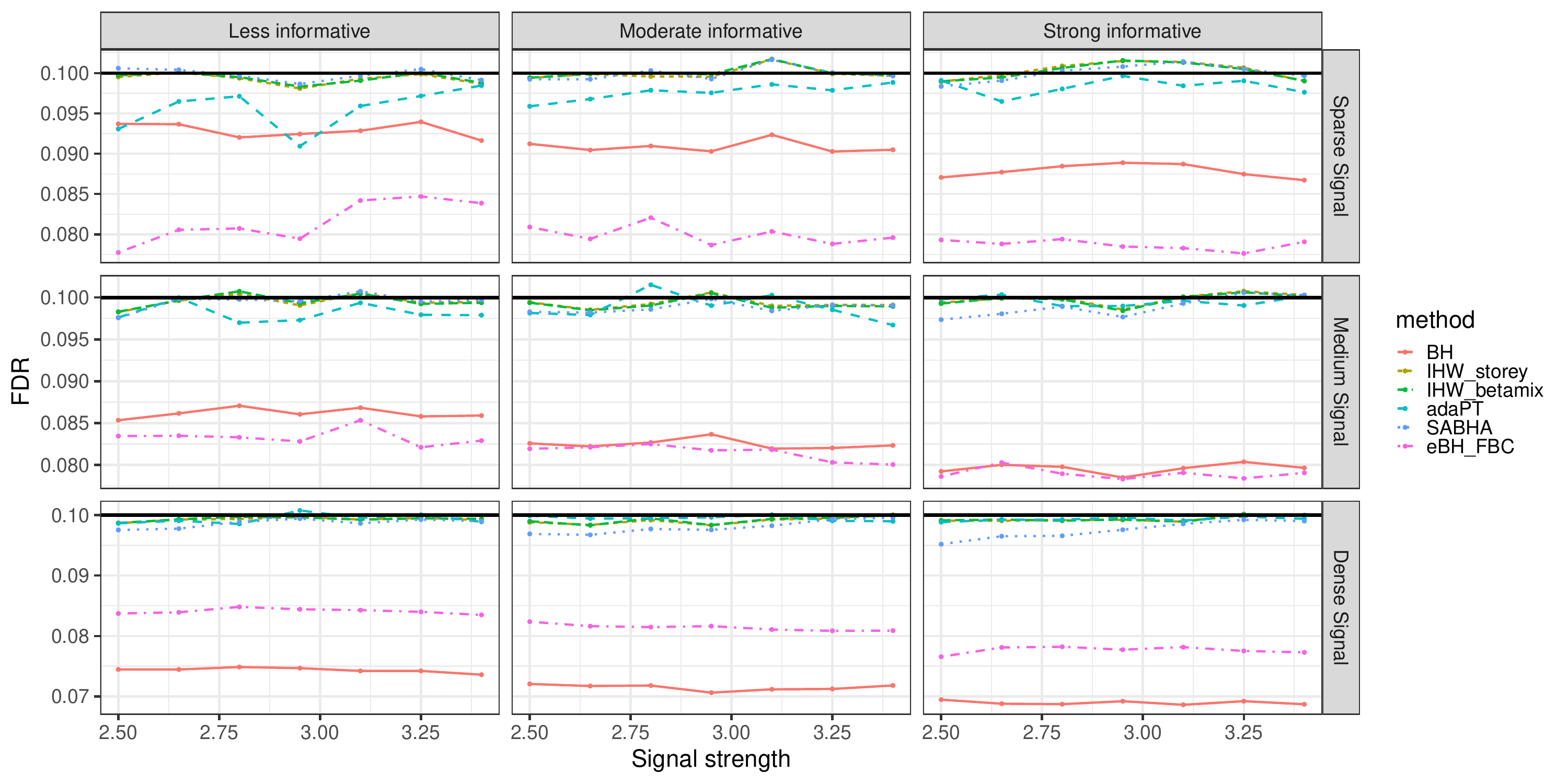}
    \caption{Empirical FDR and power for $a_f = 0.5$. Signal sparsity is controlled by setting $a_0\in\{3.5,2.5,1.5\}$, giving rise to sparse, moderate, and dense alternatives, respectively. Covariate informativeness is tuned via $a_1\in\{1.5,2,2.5\}$, corresponding to weak, moderate, and strong auxiliary signals.}
    \label{fig:ebh-gbc-af0.5}
\end{figure}

The results for $a_f = 0.5$ are presented in Figure~\ref{fig:ebh-gbc-af0.5}. When the signal is sparse and the covariate is less informative, slight FDR inflation is observed in \texttt{SABHA} and the two versions of \texttt{IHW}. \texttt{eBH\_FBC} has the highest power, followed by \texttt{SABHA} and the two versions of \texttt{IHW}. \texttt{AdaPT}, on the other hand, shows a power loss when compared to the BH procedure. However, as the covariate becomes more informative, all structure adaptive methods outperform the BH procedure, and \texttt{eBH\_FBC} has the most true discoveries when the signal is sparse. Furthermore, when the signal becomes dense, \texttt{eBH\_FBC}, \texttt{AdaPT}, and \texttt{SABHA} have similar performance in power.

\begin{figure}
    \centering
    \includegraphics[width = 0.95\textwidth]{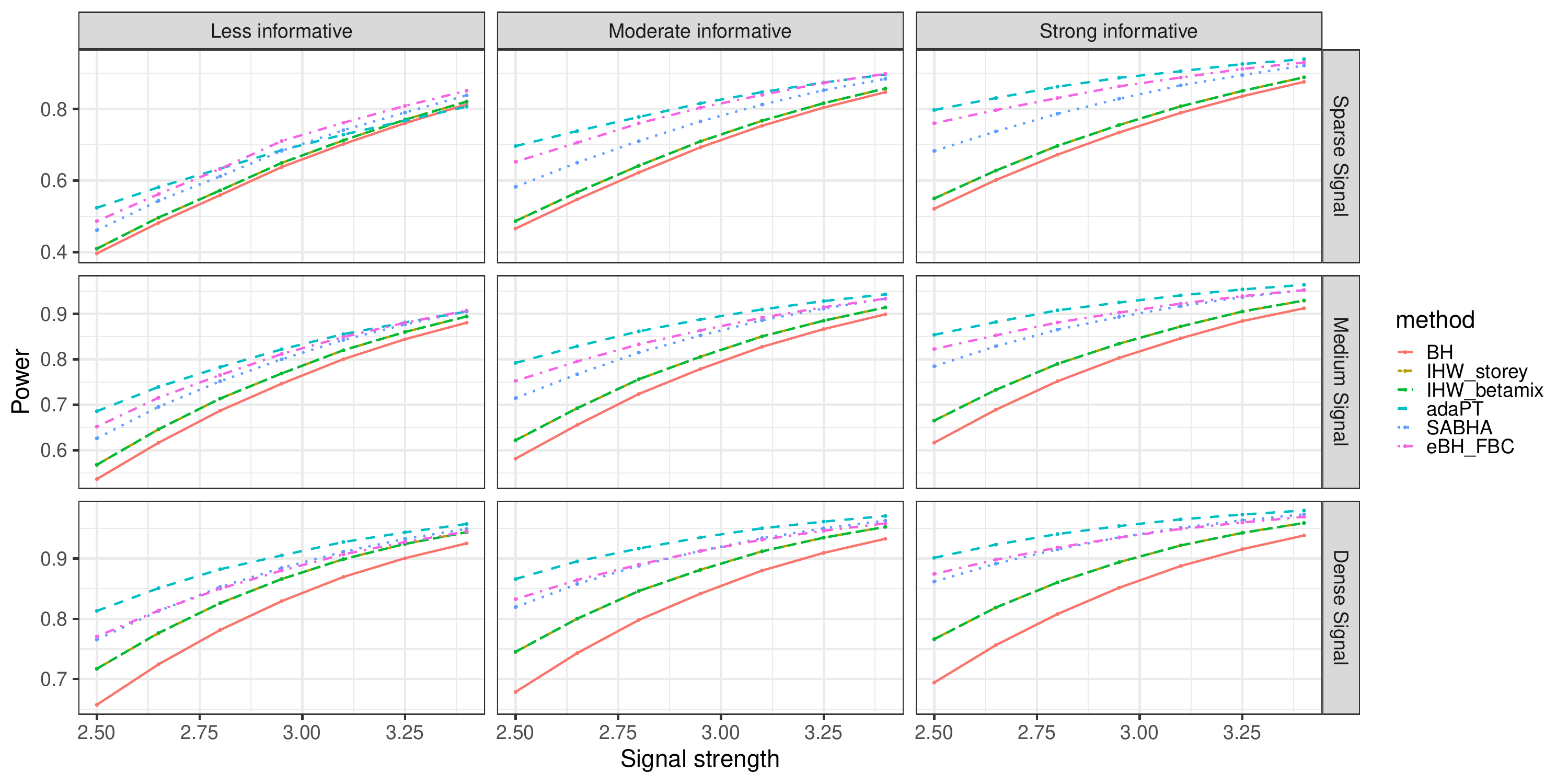}\\
    \includegraphics[width = 0.95\textwidth]{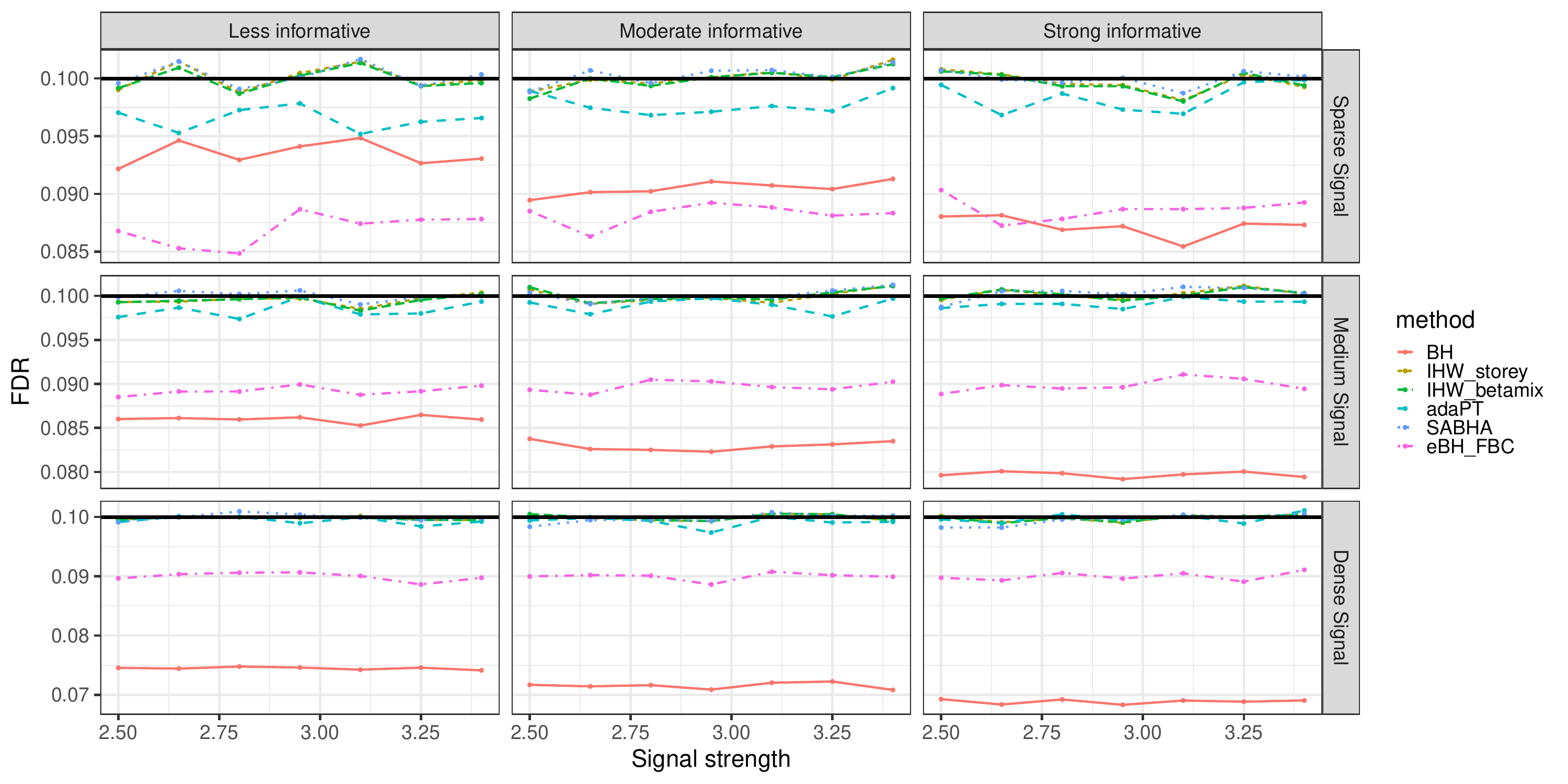}
    \caption{Empirical FDR and power for $a_f = 0$. Signal sparsity is controlled by setting $a_0\in\{3.5,2.5,1.5\}$, giving rise to sparse, moderate, and dense alternatives, respectively. Covariate informativeness is tuned via $a_1\in\{1.5,2,2.5\}$, corresponding to weak, moderate, and strong auxiliary signals.}
    \label{fig:ebh-gbc-af0}
\end{figure}

Figure~\ref{fig:ebh-gbc-af0} shows the results for $a_f = 0$, i.e., the alternative p-value distribution is independent of the covariates. In this case, \texttt{AdaPT} performs the best, followed by eBG\_FBC and \texttt{SABHA}, which dominate \texttt{IHW} and the BH procedure.

\begin{figure}
    \centering
    \includegraphics[width = 0.9\textwidth]{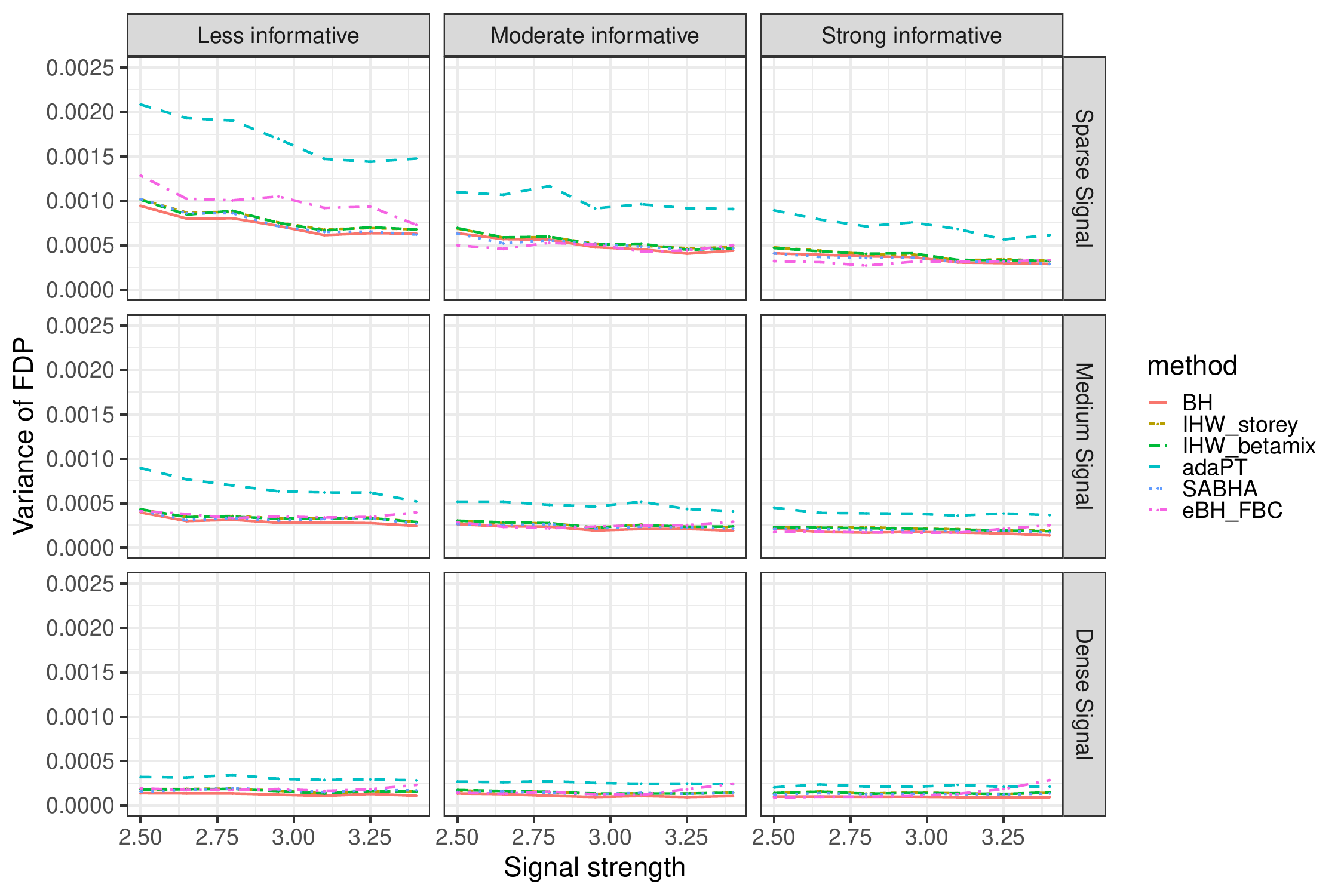}
    \caption{Variance for empirical FDR and power when $a_f = 1$. Signal sparsity is controlled by setting $a_0\in\{3.5,2.5,1.5\}$, giving rise to sparse, moderate, and dense alternatives, respectively. Covariate informativeness is tuned via $a_1\in\{1.5,2,2.5\}$, corresponding to weak, moderate, and strong auxiliary signals.}
    \label{fig:ebh-gbc-var-af1}
\end{figure}

\begin{table}
    \centering
    \begin{tabular}{|c|cccccc|}
    \hline
    method &BH&IHW\_Storey&IHW\_betamix&AdaPT&SABHA&eBH\_FBC \\
    \hline
    time in seconds  & $3\times 10^{-4}$ & $2.3\times10^{-3}$ & $1.9\times10^{-3}$&4.24&$6\times10^{-4}$&0.45\\
    \hline
    \end{tabular}
    \caption{The average running time for each method in 100 simulations. }
    \label{tab:struc-time}
\end{table}

\end{document}